\documentclass[paper=a4, 12pt]{article}
\usepackage{moreverb,url}

\usepackage[colorlinks,bookmarksopen,bookmarksnumbered,citecolor=red,urlcolor=red]{hyperref}

\usepackage[english]{babel}															
\usepackage[protrusion=true,expansion=true]{microtype}				
\usepackage{amsmath,amsfonts,amsthm, amssymb,amsbsy}										
\usepackage[pdftex]{graphicx}														
\usepackage{epstopdf}																	
\usepackage{subfig}																		
\usepackage{booktabs}																	
\usepackage{latexsym}
\usepackage{natbib}

\usepackage{authblk}
\usepackage{amsmath}
\usepackage{amsmath,amsfonts,amssymb,amsthm,epsfig,epstopdf,url,array}

\usepackage[T1]{fontenc}
\usepackage{epsfig}
\usepackage{graphicx}
\usepackage{epstopdf}
\usepackage{amssymb}
\usepackage{amsbsy}
\usepackage{multirow}
\usepackage{mathrsfs}
\usepackage{threeparttable}

\newcommand{\bs}{\boldsymbol}

\newcommand{\tup}{\textup}
\newtheorem{theorem}{Theorem}

\newcommand\BibTeX{{\rmfamily B\kern-.05em \textsc{i\kern-.025em b}\kern-.08em
T\kern-.1667em\lower.7ex\hbox{E}\kern-.125emX}}

\title{		Robust functional regression model for marginal mean and subject-specific inferences
}

\author[1]{Chunzheng Cao}
\author[2]{Jian Qing Shi}
\author[3]{Youngjo Lee 
\thanks{Corresponding author: youngjo@snu.ac.kr}}	
\smallskip
\affil[1]{School of Mathematics and Statistics, Nanjing University of Information
	Science and Technology, Nanjing, China }
\affil[2]{School of Mathematics \& Statistics, Newcastle University, UK}  
\affil[3]{Department of Statistics, Seoul National University, Seoul, Korea}   
\date{\today}

\begin{document}
\maketitle






\begin{abstract}
We introduce flexible robust functional regression models, using various
heavy-tailed processes, including a Student $t$-process. We propose
efficient algorithms in estimating parameters for the marginal mean
inferences and in predicting conditional means as well interpolation and
extrapolation for the subject-specific inferences. We develop bootstrap
prediction intervals for conditional mean curves. Numerical studies show
that the proposed model provides robust analysis against data contamination
or distribution misspecification, and the proposed prediction intervals
maintain the nominal confidence levels. A real data application is presented
as an illustrative example.
\end{abstract}

\section{Introduction}

Functional regression models are often used to analyze the medical data.
However, conclusions from functional regression models can be sensitive to
the presence of outliers or misspecifications of model assumptions.
Prediction intervals (PIs) for subjects-specific inferences have not been
well developed. In this paper, we propose robust functional regression
models and study predictions of individual response curves and their
prediction intervals.

Gaussian process (GP) has been widely used to fit an unknown regression
function to describe the nonparametric relationship between a response
variable and a set of multi-dimensional covariates.\citep{Rasmussen06}
Variety of covariance functions provide flexibility on fitting data with
different degrees of smoothness and nonlinearity. \citet{Shi07} introduced
the GP functional regression (GPFR) models, which fit the mean structure and
covariance structure simultaneously. This enables to estimate a
subject-specific regression curve consistently. The idea is further extended
to model nonlinear random-effects and is applied to construct a
patient-specific dose-response curve.\citep{Shi12} Recent applications of
GPFR models can be found in, for example, single-index model
\citep{Gramacy12} and model of non-Gaussian functional data.\citep{Wang14}

In the GPFR models, the mean structure can describe marginal mean curves
using information from all subjects, while the covariance structure can
catch up subject-specific characteristics. However, as we shall see, the
GPFR models are not robust. They are sensitive to misspecification of
distributions and the presence of outliers. In this paper, we propose to use
heavy-tailed processes (HPs) to overcome the drawbacks of GPs. Specifically
we will use scale mixtures of GP (SMGP),\citep{Rasmussen06} an extension of
scale mixtures of normal (SMN) distribution.\citep{Andrews74} The latter is
a subclass of elliptical distribution family, including Student-$t$, slash,
exponential power, contaminated-normal and other distributions. SMN
distributions have been used in various models, including nonlinear
mixed-effects models,\citep{Lachos11,Meza12} measurement error
models,\citep{Cao15,Blas16} functional models,\citep{Zhu11,Osorio16} and
extended to double hierarchical generalized linear models.\citep{Lee06}
Similarly SMGP includes many different heavy-tailed stochastic processes
such as the Student $t$-process. The SMGP inherits most of the good features
from GP; it is easy to understand its covariance structure, and it can use
many different covariance functions to allow a flexible model. SMGP includes
GP as a special case. Moreover, SMGPs except GP are HPs and they provide a
robust analysis against distributional misspecification or data
contamination.

In this paper, we extend the GPFR model \citep{Shi07,Shi12} to a HP
functional regression (HPFR) model for a robust analysis. For
maximum-likelihood estimators (MLEs), we develop an efficient EM algorithm
for proposed HPFR models. \citet{McCulloch11} investigated the impact of
distribution misspecification in linear and generalized linear models, and
showed that the overall accuracy of prediction is not heavily affected by
mild-to-moderate violations of the distribution assumptions. However, the
improvement of using HPFR models becomes significant for the data
contamination by outliers. A comprehensive simulation study is given in
Section 5.

Existing PIs for subject-specific inferences often give liberal intervals as
we shall show later. \citet{Lee16} studied PIs for random-effect models and
we extend them to HPFR models. We show that they maintain the nominal confidence
level (NCL) by using numerical studies.

The rest of this paper is organized as follows. In Section 2, we define the
HPFR model along with a brief introduction of the GPFR model. In Section 3,
we apply the HPFR model to analyze the renal anaemia data. We demonstrated
that the proposed HPFR model provides a robust analysis against outliers and
therefore avoids a misleading conclusion on a dose-response relation. In
Section 4 the estimation and prediction procedures are described and the
information consistency of subject-specific response prediction is shown. In
Section 5, a simulation study is presented to evaluate the performance of
the HPFR model and proposed procedures. Concluding remarks and discussion are given in
Section 6. All the technical details are provided as the supplementary
materials.

\section{Model}

\subsection{The GPFR model}

Consider a mixed-effects concurrent GPFR model,\citep{Shi12} defined as:
\begin{equation}  \label{model}
\begin{split}
y_{m}(t)& =\mu _{m}(t)+\tau _{m}(t)+\varepsilon _{m}(t), \\
\mu _{m}(t)& =\bs{v}_{m}^{\top }(t)\bs{\gamma}+\bs{u}_{m}^{\top }\bs{\beta}(t), \\
\tau _{m}(t)& =\bs{w}_{m}^{\top }(t)\bs{b}_{m}+\zeta _{m}(%
\bs{x}_{m}(t)),
\end{split}
\end{equation}
where $y_{m}(t)$ is a response curve for the $m$-th subject, depending on
covariates $\bs{u}_{m}$ of dimension $p_{u}$\ and functional
covariates $\bs{v}_{m}(t)$, $\bs{w}_{m}(t)$ and $%
\bs{x}_{m}(t)$ of dimensions $p_{v}$, $p_{w}$ and $p_{x}$
respectively. The model is composed of three parts: the marginal mean (related to the
so-called population-average) \citep{Diggle96}
\begin{equation*}
\tup{E}[y_{m}(t)]=\mu _{m}(t),
\end{equation*}
the random-effect $\tau _{m}(t)$ for the $m$-th subject to give the
conditional (or the so-called subject-specific) mean
\begin{equation*}
\tup{E}[y_{m}(t)|\bs{b}_m,\zeta _{m}(\bs{x}_{m}(t))]=\mu
_{m}(t)+\tau _{m}(t),
\end{equation*}
and the random error $\varepsilon _{m}(t)$. In this paper, we show how to
make marginal and conditional inferences based on functional models as above.\citep{Lee04}

In the marginal mean $\mu_m(t)$, $\bs{v}_{m}^{\top }(t)\bs{\gamma}$ is
an ordinary linear regression model with functional covariates
$\bs{v}_{m}(t)$ and regression coefficients $\bs{\gamma}$,
while $\bs{u}_{m}^{\top }\bs{\beta}(t)$ is proposed \citep{Ramsay05}
for nonparametric functional estimation using covariates
$\bs{u}_{m}$ with unknown functional coefficients $\bs{\beta}%
(t)$. Here the $p_{u}$-dimensional functional coefficient $\bs{\beta}%
(t)$ can be approximated by a set of basis functions. In this paper, we use
B-splines, i.e., $\bs{\beta}(t)=\bs{B}^{\top }%
\bs{\Phi}(t)$, in which $\bs{B}$ is a $D\times p_{u}$ matrix
with element $\beta _{ij}$, and $\bs{\Phi}(t)=(\Phi _{1}(t),\ldots
,\Phi _{D}(t))^{\top }$ are the B-spline basis functions.

In the random-effects $\tau_m(t)$, $\bs{w}_{m}^{\top }(t)\bs{b}_{m}$ is
an ordinary linear random effect model with functional covariates $%
\bs{w}_{m}(t)$ and random-effects $\bs{b}_{m}\sim \tup{N}%
_{p_{w}}(\bs{0},\bs{\Sigma}_{b})$ with, for simplicity, $%
\bs{\Sigma}_{b}$ being a diagonal matrix with elements $%
\bs{\phi}_{b}=(\phi _{1},\ldots ,\phi _{p_{w}})^{\top }$, while $%
\zeta _{m}(\bs{x}_{m}(t))$ is a functional (non-linear)
random-effects by using a GP with zero mean and covariance kernel $C(%
\bs{x}_{m}(t),\bs{x}_{m}(t^{\prime });\bs{\theta})$.
A common choice for this kernel is the following squared exponential
function
\begin{equation}
C(\bs{x}_{m}(t),\bs{x}_{m}(t^{\prime });\bs{\theta}%
)=v_{0}\exp \{-\frac{1}{2}\sum_{k=1}^{p_{x}}w_{k}(x_{m,k}(t)-x_{m,k}(t^{%
\prime }))^{2}\}.  \label{kernel}
\end{equation}
Other choices of the kernel have been studied.\citep{Rasmussen06,Shi11}
Linear random-effects can provide a clear physical explanation between the
response and the covariates, and can indicate which variables are the cause
of the variation among different subjects. The unexplained part can be
modeled by the functional random-effects.
Note that, $\mu_m(t)$ is one function describing the marginal means of all subjects,
while random functions
$\tau_m(t)$ allow different functional and nonparametric effects
for each subject to catch up subject-specific characteristics.

The random error $\varepsilon _{m}(t)$ follows $\tup{N}(0,\phi
_{\varepsilon })$ and is independent at different $t$. For convenience, we
can include it into the random-effects by
\begin{equation*}
\widetilde{\tau }_{m}(t)=\tau _{m}(t)+\varepsilon _{m}(t).
\end{equation*}
Then, $\widetilde{\tau }_{m}(t)$ is a GP with zero mean and the covariance
kernel
\begin{equation}
\begin{split}
& \widetilde{C}(\bs{x}_{m}(t),\bs{x}_{m}(t^{\prime });%
\bs{w}_{m}(t),\bs{w}_{m}(t^{\prime });\bs{\theta},%
\bs{\phi}_{b},\phi _{\varepsilon }) \\
=& \big[C(\bs{x}_{m}(t),\bs{x}_{m}(t^{\prime });%
\bs{\theta})+\sum_{k=1}^{p_{w}}\phi _{k}w_{m,k}(t)w_{m,k}(t^{\prime
})+\phi _{\varepsilon }\delta (t,t^{\prime })\big],
\end{split}
\label{covariance1}
\end{equation}
where $\delta (\cdot ,\cdot )$ is the Kronecker delta function. Thus, the
GPFR model for $y_{m}(t)$ is a GP with mean $\mu _{m}(t)$ and covariance
kernel $\widetilde{C}$.

\subsection{The HPFR model}

In this paper we show that the GPFR model provides sensitive analysis to
the presence of outliers and misspecification of distribution assumptions.
Thus, the inferences based on the GPFR model can be misleading. We show how
to make robust analysis for marginal mean and subject-specific
inferences using HPFR models.

Given a latent variable $r_{m}$, suppose that the conditional process of $%
\widetilde{\tau }_{m}(t)$ follows a GP with zero mean, but with the
covariance
\begin{equation}
\kappa (r_{m})\widetilde{C}(\bs{x}_{m}(t),\bs{x}%
_{m}(t^{\prime });\bs{w}_{m}(t),\bs{w}_{m}(t^{\prime });%
\bs{\theta},\bs{\phi}_{b},\phi _{\varepsilon }),
\label{covariance}
\end{equation}
where $\widetilde{C}(\cdot ,\cdot )$ is the covariance function given in %
\eqref{covariance1}, $\kappa (\cdot )$ is a strictly positive function, and
the latent random variable $r_{m}$ takes positive value with the cumulative
distribution function (cdf) $\tup{H}(\cdot ;\bs{\nu})$ and
probability density function (pdf) $h(\cdot ;\bs{\nu})$ with $%
\bs{\nu}$ being the degree of freedom. The property of this process
depends on the choice of $\kappa (\cdot )$ and $\tup{H}(\cdot ;%
\bs{\nu})$. When $\kappa (\cdot )\equiv 1$, it degenerates to the GP.

\citet{Lange93} studied SMN distribution with $\kappa (r_{m})=1/r_{m}$. %
\citet{Rasmussen06} called the process with conditional covariance kernel %
\eqref{covariance} a SMGP and in this paper we call it HP to highlight that
it can be applied to wider class of data using double hierarchical
generalized linear models.\citep{Lee06} If $r\sim \tup{Gamma}(\nu
/2,\nu /2)$, it becomes a Student-$t$ (T) process, if $r\sim \tup{Beta}%
(\nu ,1)$, it becomes a slash (SL) process and if $P(r=\gamma )=\nu
,~P(r=1)=1-\nu $ with $0<\nu \leqslant 1,0<\gamma \leqslant 1$, it is a
contaminated-normal (CN) process. We call model \eqref{model} under this HP
structure a mixed-effects HPFR model (abbreviated as `HPFR'). HPFR models
can cover existing heavy-tailed mixed-effects models,
\citep{Pinheiro01,Savalli06} a robust P-splines model \citep{Osorio16} and
extended T-process regression models,\citep{Wang16} etc.

\section{An illustrative example: the renal anaemia data}

\citet{West07} studied the renal anaemia data, which contain 74 dialysis
patients who received Darbepoetin alfa (DA) to treat their renal anaemia.
The Hemoglobin (Hb) levels and epoetin dose were recorded from the original
study period of 9 months with a further 3 months extension. The doses of
epoetin DA were determined by a strict clinical decision support system %
\citep{Tolman05,Will07} to maintain the Hb level around 11.8 g/dl, as a
target for dialysis patients. The lower Hb level will lead the patient to
anemia, while the higher level increases the risk of prothrombotic and other
problems. The experiment is typically a dose-response study to evaluate the
control of Hb levels with the agent DA. The GPFR model has been used to
analyze the data.\citep{Shi12} The response is the Hb level, and the dose
of DA and time are considered as predictors.

Figure~1(1) displays the Hb levels measured for each patient
in a period of 12 months, and Figure~1(3) gives the histogram
of dosages of DA for all patients. Since the distribution of dosage $%
dose_{m}(t)$ (for the $m$-th patient at the month $t$)
is quite skew, we use a log-transformation $\log [10dose_{m}(t)+1]$
to reduce the impact of extreme values. As we see in
Figure~1(4), the converted dose values except zeros are
nearly normal after using the transformation. Taking into account a certain
lag-time of the drug reaction, \citet{Shi12} choose $dose_{m}(t-2)$, i.e.,
the dosage of DA taken two months before, as a key covariate. Following %
\citet{Shi12}, we use a linear regression model for the marginal mean
\begin{equation*}
\mu _{m}(t)=\gamma _{0}+\gamma _{1}\cdot t+\gamma _{2}\cdot
dose_{m}(t-2)+\gamma _{3}\cdot dose_{m}(t-2)^{2},
\end{equation*}
and random-effects
\begin{equation*}
\tau _{m}(t)=b_{m}+\zeta _{m}(\bs{x}_{m}(t)),
\end{equation*}
with $\bs{x}_{m}(t)=[t,dose_{m}(t-2)]^{\top }$. Figure~1(2) shows
the index plots of Mahalanobis distance (defined later) of each patient
under the GPFR model, which follows theoretically a $\chi ^{2}$%
-distribution. We can see that there are large Mahalanobis distance values
from the four patients beyond the quantile $\chi _{0.01}^{2}$ cutoff line.
These four patients are recognized as potential outliers.

\begin{figure}[h!]
	\centering
	\fbox{%
		\includegraphics[width=9cm,height=6cm]{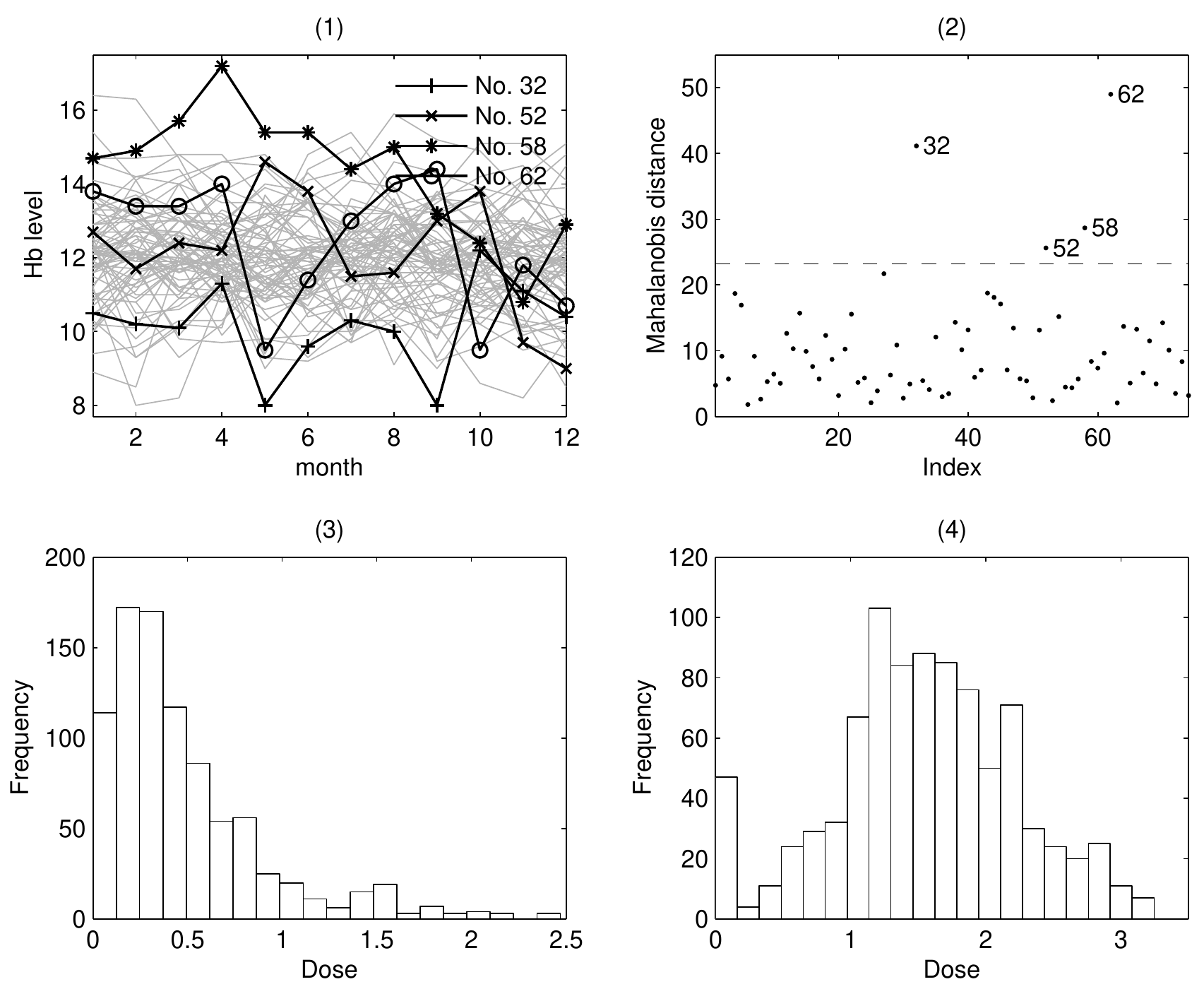} }
	\caption{{\protect\scriptsize Renal data: (1) the Hb levels for the 74
			patients; (2) Mahalanobis distance under the GPFR model; (3) Histogram of
			dose values before transformation; (4) Histogram of dose values after
			transformation.}}
	\label{exam-fig-1}
\end{figure}

\subsection{The population-average inferences}

Table~\ref{exam-tab-1} reports the MLEs of coefficients for the marginal
mean and their standard errors (the first value in parentheses). Here `N'
stands for the GPFR model. The degree parameters for three heavy-tailed
models are estimated together with other parameters. In Table~\ref
{exam-tab-1}, the negative coefficient of $t$ means that the patients have a
decreasing trend in the level of Hb over time without considering dose
effects. Note that the values of Bayesian information criterion (BIC) %
\citep{Schwarz78} and the values of standard errors (the second values of each covariate in fixed-effects) for three heavy-tailed
models are often smaller than those for the GPFR model, and the model using
T-process (TPFR) is the best choice under BIC. We also report the values of
root of mean squared error (RMSE) of the predictions for all patients in
Table~\ref{exam-tab-1}. The smaller values of RMSE under three HPFR models
confirm the better performance of the heavy-tailed models in fitting the
data. The third values of each covariate in fixed-effects in Table~\ref{exam-tab-1} are
relative change ratios (\%) of the estimates after removing the data from
four potential outlier patients. We find that the change ratios of estimates
under three HPFR models are much smaller than those under the GPFR model,
which indicate the robustness in parameter estimation by using the
heavy-tailed models.

\setlength{\tabcolsep}{0.25em}
\begin{table}[h]
\scriptsize\sf\centering
\begin{threeparttable}[b]
\caption{Parameter estimation of the renal anaemia data.\label{exam-tab-1}}
\begin{tabular}{cccccccc}
\toprule
\multirow{2}{*}{Model} & \multirow{2}{*}{Degree} & \multirow{2}{*}{BIC} & %
\multirow{2}{*}{RMSE} & \multicolumn{4}{c}{Covariates in fixed-effects} \\
\cline{5-8}
&  &  &  & Constant & $t$ & $dose_m(t-2)$ & $dose_m^2(t-2)$ \\
\midrule
N & / & 2136 & 0.416 & 11.500/0.310/\textbf{-1.4} & -0.036/0.022/\textbf{-32.9} & 0.863/0.313/\textbf{-6.1} & -0.196/0.103/\textbf{-22.1} \\
T & {\tiny $\widehat{\nu}=7.106$} & 2088 & 0.393 & 11.426/0.297/-0.4 & -0.035/0.021/-14.7 & 0.749/0.304/-0.6 & -0.097/0.108/~~2.8 \\
SL & {\tiny $\widehat{\nu}=1.723$} & 2092 & 0.391 & 11.451/0.299/-0.5 & -0.035/0.021/-15.2 & 0.746/0.308/~0.3 & -0.106/0.109/~~2.9 \\
CN & {\tiny $
\begin{array}{c}
\widehat{\nu}=0.424 \\
\widehat{\gamma}=0.326
\end{array}
$} & 2101 & 0.392 & 11.464/0.285/-0.6 & -0.040/0.020/-17.8 & 0.744/0.267/-3.3 & -0.094/0.099/-11.7 \\
\bottomrule
\end{tabular}
\begin{tablenotes}
\item RMSE: root of mean squared error of the predictions;
\item Three values of each covariate in fixed-effects are in turn, the values of estimate, standard error and the relative change ratio of the estimate after removing the outliers.
\end{tablenotes}
\end{threeparttable}
\end{table}

We plot the marginal mean curve of Hb levels affected by dose in Figure~2
under four models. The solid lines and the dash lines are drawn
respectively, based on the complete 74 patients data, and data deleting
those from the four patients. For the GPFR model, the level of Hb increases
when the dose is less than 0.8 and decreases slowly after it. However, this
turning point changes from 0.8 to 1.3 after deleting the data from the four
patients. Under all the HPFR models, the marginal mean maintains to increase
on the whole interval but the increasing rate reduces slowly at large
dosage. This is a more realistic dose-response curve based on the current
understanding on the dose-response relation. Compared with GPFR, the curve
shapes under three HPFR models have smaller changes after deleting outliers,
which implies the robustness of heavy-tailed models for population-average
inferences.

\begin{figure}[h!]
	\centering
	\fbox{%
		\includegraphics[width=9cm,height=6cm]{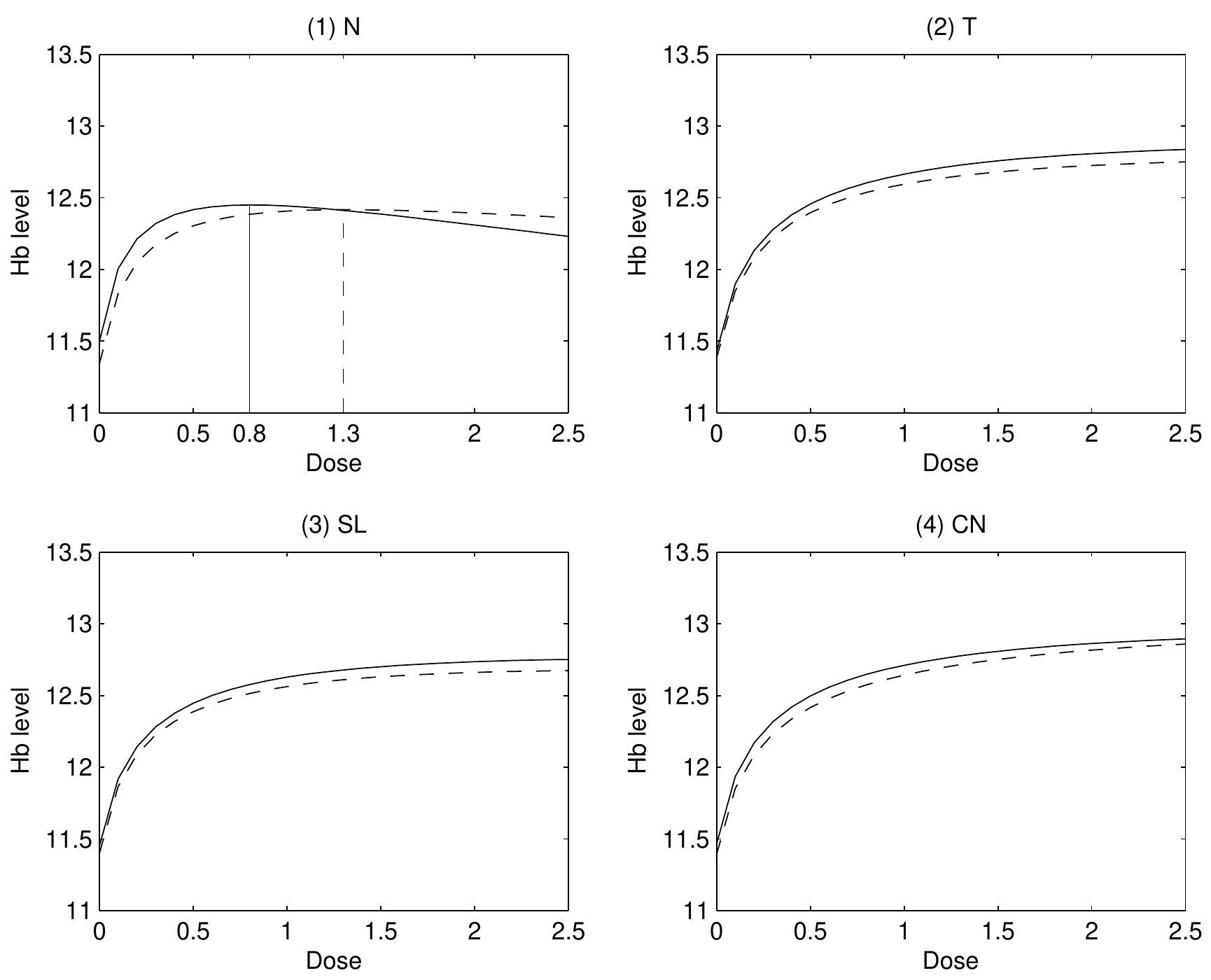} }
	\caption{{\protect\scriptsize Mean curve of dose response based on original
			data (Solid line), and data without outliers (Dash line).}}
	\label{exam-fig-2}
\end{figure}

\subsection{The subject-specific inferences}

We now study the patient-specific dose-response curves. We use the data from
the first 12 months, and predict (extrapolate) the Hb level at the 14-th
month for the dosage from 0 to 2.5. Figure~3 shows the
predictions of the patient-specific dose-response curves under the GPFR and
TPFR model for two typical patients. For the first patient, the
dose-response curve under the GPFR model does not seem realistic, because it
cannot achieve the target value of 11.8 even if we increase the dosage.
However, the TPFR curve reaches the target when dosage is increased to 1.75.
This is the dosage the patient should take at Month 12 if they wants to
maintain the Hb level around 11.8 at Month 14. For the second patient, the
dose-response curve suggests that the Hb level under the GPFR model reaches
the target when dose is 1, and it stays at the same level even when the
dosage increases. The TPFR model achieves a more reasonable dose-response
relation: the Hb level reaches the target when dose is 0.75 and keep
increasing as the dosage increases. We also plotted the PIs for
patient-specific curves. Note that the narrowest interval is allocated to
the Hb level around the actual dosages the patients took in the previous
month. This is because the covariance kernel we used in the stochastic
process depends on the difference of dosages between consecutive months, and
thus the prediction has less uncertainty in the neighbour of the previous
dosage. We will show later how to construct the PIs and show that the
proposed PIs maintain the NCLs.

\begin{figure}[h!]
	\centering
	\fbox{%
		\includegraphics[width=9cm,height=6cm]{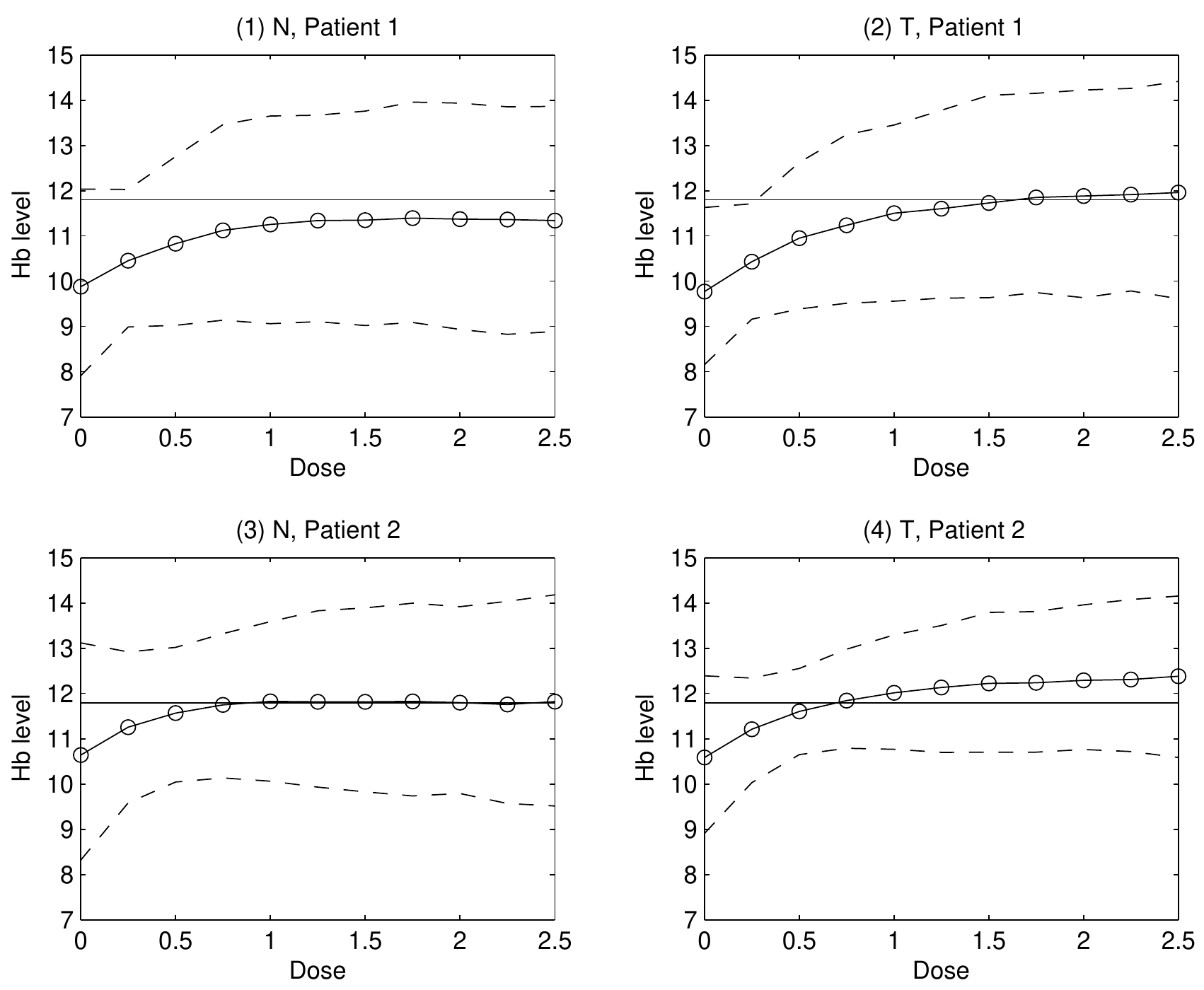}
	}
	\caption{{\protect\scriptsize Dose-response of patient-specific prediction
			for the first selected patient (the upper panel) and the second patient (the
			lower panel) under the GPFR (N) and TPFR (T) model. The solid lines stand
			for predictions, and the dash lines are 95\% predictive bands.}}
	\label{exam-fig-3}
\end{figure}

\section{Inference methods}

Suppose that we have data from $M$ different subjects, and all functional
covariates are observed at points $\bs{t}_{m}=(t_{m1},\ldots
,t_{m,n_{m}})^{\top }$ for the $m$-th subject. Let $\bs{y}%
_{m}=(y_{m1},\ldots ,y_{m,n_{m}})^{\top }$, $\bs{V}_{m}$ be the $%
n_{m}\times p_{v}$ matrix with the $i$-th row $\bs{v}_{m}^{\top
}(t_{mi})$, $\bs{X}_{m}$ and $\bs{W}_{m}$ be the matrix
defined in the same way as $\bs{V}_{m}$. Consider the following HPFR
model
\begin{equation}
\bs{y}_{m}=\bs{\Phi}_{m}\bs{B}\bs{u}_{m}+%
\bs{V}_{m}\bs{\gamma}+\bs{W}_{m}\bs{b}_{m}+%
\bs{\zeta}_{m}+\bs{\varepsilon}_{m},  \label{HPFR}
\end{equation}
where $\bs{\Phi}_{m}$ is an $n_{m}\times D$ matrix with the $i$-th
row $\bs{\Phi}^{\top }(t_{mi})$, and
\begin{equation}
\begin{split}
\bs{\varepsilon}_{m}|r_{m}& \overset{ind}{\sim }\tup{N}_{n_{m}}(%
\bs{0},\kappa (r_{m})\phi _{\varepsilon }\bs{I}_{n_{m}}), \\
\bs{b}_{m}|r_{m}& \overset{ind}{\sim }\tup{N}_{p_{w}}(%
\bs{0},\kappa (r_{m})\bs{\Sigma}_{b}), \\
\bs{\zeta}_{m}|r_{m}& \overset{ind}{\sim }\tup{N}_{n_{m}}(%
\bs{0},\kappa (r_{m})\bs{C}_{m}), \\
r_{m}& \overset{iid}{\sim }\tup{H}(r;\bs{\nu}),m=1,\ldots ,M,
\end{split}
\end{equation}
$\bs{C}_{m}$ is an $n_{m}\times n_{m}$ covariance matrix with the $%
(i,j)$-th element $C(\bs{x}_{m}(t_{i}),\bs{x}_{m}(t_{j});%
\bs{\theta})$. Then, $\bs{y}_{m}$ follows a SMN distribution,
\citep{Andrews74,Fang90} i.e., $\tup{SMN}(\bs{\mu}_{m},%
\bs{\Sigma}_{m};\tup{H})$, where $\bs{\mu}_{m}=%
\bs{\Phi}_{m}\bs{B}\bs{u}_{m}+\bs{V}_{m}%
\bs{\gamma}$ and $\bs{\Sigma}_{m}=\bs{C}_{m}+%
\bs{W}_{m}\bs{\Sigma}_{b}\bs{W}_{m}^{\top }+\phi
_{\varepsilon }\bs{I}_{n_{m}}$.

\subsection{Parameter estimation}

Denote the parameter set $\{\bs{B},\bs{\gamma},%
\bs{\theta},\bs{\phi}_{b},\phi _{\varepsilon },%
\bs{\nu}\}$ by $\bs{\Theta}$. Let $\bs{\mathcal{D}}%
_{c}=\{\bs{\mathcal{D}}_{m},r_{m},m=1,\ldots ,M\}$ be the complete
data, where $\bs{\mathcal{D}}_{m}=\{\bs{y}_{m},\bs{u}%
_{m},\bs{V}_{m},\bs{W}_{m},\bs{X}_{m},\bs{t}%
_{m}\}$ be the observed data of the $m$-th subject. For convenience, we
separate $\bs{\Theta}$ into $\bs{\beta}=(\tup{Vec}(%
\bs{B})^{\top },\bs{\gamma}^{\top })^{\top }$, $%
\bs{\psi}=(\bs{\theta}^{\top },\bs{\phi}_{b}^{\top
},\phi _{\varepsilon })^{\top }$ and $\bs{\nu}$, where $%
\bs{\nu}$ is a parameter for the degree of freedom.

The log-likelihood of the complete data is given by
\begin{equation}  \label{comp-loglihood}
\begin{split}
l(\bs{\Theta}|\bs{\mathcal{D}}_c)=&-\frac{1}{2}\sum_{m=1}^M
\log |\bs{\Sigma}_m|-\frac{1}{2}\sum_{m=1}^M n_m\log
\kappa(r_m)+\sum_{m=1}^M \log h(r_m;\bs{\nu}) \\
&-\frac{1}{2}\sum_{m=1}^M \kappa^{-1}(r_m)(\bs{y}_m-\bs{A}_m%
\bs{\beta})^{\top}\bs{\Sigma}_m^{-1}(\bs{y}_m-%
\bs{A}_m\bs{\beta}),
\end{split}
\end{equation}
where $\bs{A}_m=[\bs{u}_m^{\top}\otimes \bs{\Phi}_m, %
\bs{V}_m]$ satisfies $\bs{\mu}_m=\bs{A}_m%
\bs{\beta}$. For the MLEs of $\bs{\Theta}$, we propose to
use an EM algorithm below.

E-step: Given the current value of $\bs{\Theta}^{(k)}$, we calculate
the $Q$-function\newline
$\tup{E}\big[l(\bs{\Theta}|\bs{\mathcal{D}}_{c})\big|%
\bs{\Theta}^{(k)},\bs{\mathcal{D}}_{m},m=1,\ldots ,M\big]$
which is proportional to
\begin{equation}
Q(\bs{\Theta}|\bs{\Theta}^{(k)})=Q_{1}(\bs{\beta},%
\bs{\psi}|\bs{\Theta}^{(k)})+Q_{2}(\bs{\nu}|%
\bs{\Theta}^{(k)}),  \label{Qfun}
\end{equation}
with
\begin{align}
Q_{1}(\bs{\beta},\bs{\psi}|\bs{\Theta}^{(k)})=& -%
\frac{1}{2}\sum_{m=1}^{M}\log |\bs{\Sigma}_{m}|-\frac{1}{2}%
\sum_{m=1}^{M}\pi _{m}^{(k)}(\bs{y}_{m}-\bs{A}_{m}%
\bs{\beta})^{\top }\bs{\Sigma}_{m}^{-1}(\bs{y}_{m}-%
\bs{A}_{m}\bs{\beta}),  \label{Qfun1} \\
Q_{2}(\bs{\nu}|\bs{\Theta}^{(k)})=& \sum_{m=1}^{M}\tup{E}%
[\log h(r_{m};\bs{\nu})|\bs{\Theta}^{(k)},%
\bs{\mathcal{D}}_{m}],  \label{Qfun2}
\end{align}
where
\begin{equation}
\pi _{m}^{(k)}=\tup{E}[\kappa ^{-1}(r_{m})|\bs{\Theta}^{(k)},%
\bs{\mathcal{D}}_{m}].  \label{E-step}
\end{equation}
In the HPs, the weight $\pi _{m}$ is inversely proportional to the
Mahalanobis distance $d_{m}=(\bs{y}_{m}-\bs{\mu}_{m})^{\top }%
\bs{\Sigma}_{m}^{-1}(\bs{y}_{m}-\bs{\mu}_{m})$ (see
supplementary materials). In the presence of outliers or model
misspecification, $d_{m}$ increases, leading to smaller weight to give
robust analysis.

The likelihood \eqref{comp-loglihood} is the h-likelihood of \citet{Lee96}. We can estimate the parameters by
using the h-likelihood method. Given $\bs{\Theta}$, maximizing the conditional likelihood $p_{\bs{\Theta}}(r_m|\bs{y}_m)$ with respect to $r_m$ is equivalent to maximizing the h-likelihood $L(\bs{\Theta}, r_m; \bs{y}_m, r_m)$, which has an explicit form as \eqref{comp-loglihood}. To obtain MLEs of $\bs{\Theta}$ in h-likelihood approach, the Laplace approximation has been proposed.\citep{Lee06} However, the EM algorithm is more efficient in the HPFR model, because the E-step \eqref{E-step} is straightforward.
Following ECM \citep{Meng93} and ECME,\citep{Liu94} we implement the EM algorithm as follows.

CM-step: We maximize \eqref{Qfun} with respect to $\bs{\Theta}$ to
get the updated parameter given $\bs{\Theta}^{(k)}$. We propose the
following sub-iterative process:

\begin{enumerate}
\item  Set the least square estimation of $\bs{\beta}$ as the
initial value $\bs{\beta}^{(0)}$ under the assumption: $%
\bs{\Sigma}_{m}=\bs{I}_{m}$ and $\pi _{m}=1$ for $m=1,\ldots
,M$;

\item  Obtain the initial value $\bs{\psi}^{(0)}$ by maximizing the $%
Q$-function in equation \eqref{Qfun1} under $\bs{\beta}^{(0)}$ and $%
\pi _{m}=1$;

\item  Choose a sensible initial value of $\bs{\nu}^{(0)}$;

\item  Calculate the weights $\pi _{m}$ $(m=1,\ldots ,M)$ under the current
values $\bs{\beta}$, $\bs{\psi}$ and $\bs{\nu}$;

\item  Update $\bs{\beta}$ under the current values of $\pi _{m}$, $%
\bs{\psi}$ by
\begin{equation}
\bs{\beta}^{(k+1)}=\big\{\sum_{m=1}^{M}\pi _{m}^{(k)}\bs{A}%
_{m}^{\top }\bs{\Sigma}_{m}^{-1}\bs{A}_{m}\big\}^{-1}\big\{%
\sum_{m=1}^{M}\pi _{m}^{(k)}\bs{A}_{m}^{\top }\bs{\Sigma}%
_{m}^{-1}\bs{y}_{m}\big\}\big|_{\bs{\psi}^{(k)}};
\label{mlebeta}
\end{equation}

\item  Update $\bs{\psi}$ by maximizing the $Q$-function in equation %
\eqref{Qfun1} under the current values of $\bs{\beta}$ and $\pi _{m}$%
;

\item  Given the current values of $\bs{\beta}$ and $%
\bs{\psi}$, update $\bs{\nu}$ by maximizing the constrained
actual marginal log-likelihood function \citep{Liu94} of $\{\bs{y}%
_{m},m=1,\ldots ,M\}$ over $\bs{\nu}$.
\end{enumerate}

Ideally we need to repeat steps 4 to 7 until it reaches convergence.
Practically we can set a threshold and stop the sub-iteration when $\Vert %
\bs{\Theta}^{(k+1)}-\bs{\Theta}^{(k)}\Vert $ is smaller than
the threshold. Step 6 is analogous to that under GP assumptions by
transforming the residual $\bs{e}_{m}=\bs{y}_{m}-%
\bs{A}_{m}\bs{\beta}$ into $\widetilde{\bs{e}}_{m}=%
\sqrt{\pi _{m}}(\bs{y}_{m}-\bs{A}_{m}\bs{\beta})$.
Following the ECME algorithm \citep{Liu94}, step 7 does not maximize the
expected log-likelihood \eqref{Qfun2} but maximizes the actual
log-likelihood over $\bs{\nu}$, which yields a much faster
converging speed. The actual log-likelihood of $\{\bs{y}%
_{m},m=1,\ldots ,M\}$ as well as the score function of $\bs{\Theta}$
under some SMN distributions are given in supplementary materials.

The asymptotic confidence intervals of the MLEs for this model can be
obtained through the observed information matrix $\mathbf{J}(\widehat{%
\bs{\Theta}})$ or the expected information matrix $\mathbf{I}(%
\widehat{\bs{\Theta}})$ (see supplementary materials). Here we
estimate the degree $\bs{\nu}$ together with $\bs{\beta}$
and $\bs{\psi}$. When sample size is small, the estimation of $%
\bs{\nu}$ may be not reliable. In this case, we may fix its value,
e.g., choose $\nu =4$ for Student-$t$,\citep{Lange89} or choose $%
\bs{\nu}$ by BIC or by cross-validation.

\subsection{Prediction}

Suppose that we want to predict the responses $\bs{y}^{\ast
}=(y_{1}^{\ast },\ldots ,y_{n}^{\ast })^{\top }$ of a new $(M+1)$-th subject
at points $\bs{t}^{\ast }=(t_{1}^{\ast },\ldots ,t_{n}^{\ast
})^{\top }$. Besides the data of $M$ subjects, suppose that we have $n_{M+1}$
observations for the new $(M+1)$-th subject at points $\bs{t}%
_{M+1}=(t_{M+1,1},\ldots ,t_{M+1,n_{M+1}})^{\top }$. Let $%
\bs{\mathcal{D}}_{M+1}=\{\bs{y}_{M+1},\bs{u}_{M+1},%
\bs{V}_{M+1},\bs{W}_{M+1},\bs{X}_{M+1},\bs{t}%
_{M+1}\}$ and $\bs{\mathcal{D}}=\{\bs{\mathcal{D}}%
_{m}|m=1,\ldots ,M+1\}$ be the observed data.

We rewrite model \eqref{HPFR} as
\begin{equation}
\bs{y}_{m}=\bs{\mu}_{m}+\widetilde{\bs{\tau}}_{m},
\end{equation}
with $\widetilde{\bs{\tau}}_{m}\overset{ind}{\sim }\tup{SMN}(%
\bs{0},\bs{\Sigma}_{m};\tup{H})$ for $m=1,\ldots ,M+1$.
For the $(M+1)$-th subject, let $\widetilde{\bs{\tau}}^{\ast }$ be
the random-effects at unobserved points $\bs{t}^{\ast }$, while $%
\widetilde{\bs{\tau}}_{M+1}$ be the random-effects at observed
points $\bs{t}_{M+1}$. Thus, we have
\begin{equation}
{\big(\widetilde{\bs{\tau}}_{M+1}^{\top },\widetilde{%
\bs{\tau}}^{\ast \top }\big)}^{\top }{\big|}r_{M+1}\sim \tup{N}%
_{n_{M+1}+n}(\bs{0},\kappa (r_{M+1})\widetilde{\bs{\Sigma}}%
_{M+1}),
\end{equation}
where
\begin{equation}
\widetilde{\bs{\Sigma}}_{M+1}=\left(
\begin{array}{cc}
\bs{\Sigma}_{M+1} & \bs{\Sigma}_{M+1}^{\ast } \\
\bs{\Sigma}_{M+1}^{\ast \top } & \bs{\Sigma}^{\ast }
\end{array}
\right) ,
\end{equation}
with $\bs{\Sigma}^{\ast }$ and $\bs{\Sigma}_{M+1}^{\ast }$
are, respectively, the covariance matrix of the new subject with elements
evaluated from the covariance function \eqref{kernel} at point pairs $%
(t_{j}^{\ast },t_{k}^{\ast })$ and $(t_{i},t_{k}^{\ast })$ for $i\in
\{1,\ldots ,n_{M+1}\}$ and $j,k\in \{1,\ldots ,n\}$. Then, the conditional
expectation of $\bs{y}^{\ast }$ given $\bs{\mathcal{D}}%
_{M+1} $ and $\bs{\Theta}$ is
\begin{equation}  \label{conditional expect}
\begin{split}
\tup{E}(\bs{y}^{\ast }|\bs{\mathcal{D}}_{M+1};%
\bs{\Theta})& = \tup{E}(\bs{\mu}^{\ast }+\widetilde{%
\bs{\tau}}^{\ast }|\bs{\mathcal{D}}_{M+1};\bs{\Theta}%
) \\
& = \bs{\mu}^{\ast } + \tup{E}\big[\tup{E}(\widetilde{%
\bs{\tau}}^{\ast }|r_{M+1},\bs{\mathcal{D}}_{M+1};%
\bs{\Theta})\big] \\
& =\bs{\mu}^{\ast }+\bs{\Sigma}_{M+1}^{\ast \top }%
\bs{\Sigma}_{M+1}^{-1}(\bs{y}_{M+1}-\bs{\mu}_{M+1}),
\end{split}
\end{equation}
where $\bs{\mu}^{\ast }$ and $\bs{\mu}_{M+1}$ are
respectively the marginal mean curve of the new subject at points $%
\bs{t}^{\ast }$ and $\bs{t}_{M+1}$.

The conditional variance of $\bs{y}^{\ast }$ given $\bs{%
\mathcal{D}}_{M+1}$ and $\bs{\Theta}$ is given by
\begin{equation}  \label{conditional variance}
\begin{split}
\tup{Var}(\bs{y}^{\ast }|\bs{\mathcal{D}}_{M+1};%
\bs{\Theta})&=\tup{E}(\bs{y}^{\ast }\bs{y}^{\ast \top}|%
\bs{\mathcal{D}}_{M+1};\bs{\Theta}) -\tup{E}(%
\bs{y}^{\ast }|\bs{\mathcal{D}}_{M+1};\bs{\Theta})\tup{%
E}(\bs{y}^{\ast \top}|\bs{\mathcal{D}}_{M+1};\bs{\Theta})
\\
&=\tup{E}\big[\tup{Var}(\bs{y}^{\ast }|r_{M+1},\bs{%
\mathcal{D}}_{M+1};\bs{\Theta})\big] \\
&=\tup{E}[\kappa(r_{M+1})|\bs{\mathcal{D}}_{M+1};%
\bs{\Theta}](\bs{\Sigma}^{\ast }-\bs{\Sigma}_{M+1}^{\ast \top}%
\bs{\Sigma}_{M+1}^{-1}\bs{\Sigma}_{M+1}^{\ast }).
\end{split}
\end{equation}

For GP models, PIs have been studied, assuming normal distribution with the
first two moments \eqref{conditional expect} and
\eqref{conditional
variance}.\citep{Wang14} But, the predictive distribution for the
conditional mean $p(\bs{y}^{\ast }|\bs{\mathcal{D}}_{M+1};%
\widehat{\bs{\Theta}})$ under HPFR models may not be normal. For
instance the T-process with low degree of freedom, the normal approximation
may not be appropriate. Thus, we should compute the quantiles based on the
true predictive distribution for a better prediction. However, the resulting
PIs would be liberal because they do not take account for the uncertainty in
estimation the parameters.\citep{Bjornstad90,Bjornstad96}

We propose the following parametric bootstrap method based on a finite
sample adjustment to remedy the drawback:

\begin{enumerate}
\item  Generate $\bs{\Theta}_{j}^{\ast }$ $(j=1,\ldots ,J)$ from its
asymptotic distribution $\tup{N}(\widehat{\bs{\Theta}},\mathbf{J}%
^{-1}(\widehat{\bs{\Theta}}))$;

\item  Approximate the predictive distribution of $\bs{y}^{\ast }$
by
\begin{equation}
\begin{split}
p(\bs{y}^{\ast }|\bs{\mathcal{D}})& =\int p(\bs{y}%
^{\ast }|\bs{\mathcal{D}}_{M+1};\widehat{\bs{\Theta}})p(%
\widehat{\bs{\Theta}}|\bs{\mathcal{D}})\tup{d}\widehat{%
\bs{\Theta}} \\
& \approx \frac{1}{J}\sum_{j=1}^{J}p(\bs{y}^{\ast }|%
\bs{\mathcal{D}}_{M+1};\bs{\Theta}_{j}^{\ast }).
\end{split}
\label{predictive distribution}
\end{equation}
\end{enumerate}

Note that Step (2) involves the generation of data from $p(\bs{y}%
^{\ast }|\bs{\mathcal{D}}_{M+1};\bs{\Theta}_{j}^{\ast })$
which may be not straightforward in some cases. This problem can be
addressed by augmenting the latent variable $r_{M+1}$ in the process. We
first generate $r_{M+1}$ from the conditional distribution $p(r_{M+1}|%
\bs{\mathcal{D}}_{M+1};\bs{\Theta}_{j}^{\ast })$ which is
given in supplementary materials for some distributions. Given $%
\bs{\mathcal{D}}_{M+1}$ and $r_{M+1}$, it is straightforward to
generate $\bs{y}^{\ast }$ since its conditional distribution is
Gaussian:
\begin{equation}
\bs{y}^{\ast }|r_{M+1},\bs{\mathcal{D}}_{M+1},%
\bs{\Theta}_{j}^{\ast }\sim \tup{N}_{n}(\widetilde{%
\bs{\mu}}^{\ast },\kappa (r_{M+1})\widetilde{\bs{\Sigma}}%
^{\ast }),
\end{equation}
with $\widetilde{\bs{\mu}}^{\ast }=\bs{\mu}^{\ast }+%
\bs{\Sigma}_{M+1}^{\ast \top }\bs{\Sigma}_{M+1}^{-1}(%
\bs{y}_{M+1}-\bs{\mu}_{M+1})$ and $\widetilde{%
\bs{\Sigma}}^{\ast }=\bs{\Sigma}^{\ast }-\bs{\Sigma}%
_{M+1}^{\ast \top }\bs{\Sigma}_{M+1}^{-1}\bs{\Sigma}%
_{M+1}^{\ast }$ evaluated at $\bs{\Theta}_{j}^{\ast }$. Hence, we
can obtain the prediction as well as pointwise PIs of $\bs{y}^{\ast
} $ using the sample quantiles calculated from bootstrap replications.

The bootstrap method can also be used to calculate other quantities, such as
the subject-specific random-effects. It can be considered as an extension of
PIs \citep{Lee16} for random-effects to curve predictions.

\subsection{Information consistency}

\citet{Seeger08} proved information consistency that the prediction based on
GP model can be consistent to the true curve, which has been extended to
generalized GPFR model \citep{Wang14} and T-process regression model.
\citep{Wang16} We now discuss this information consistency for the HPFR
model.

For convenience, we omit the subscript $m$ here and denote the data by $%
\bs{y}_n=(y_1,\ldots,y_n)^{\top}$ at points $\bs{t_n}%
=(t_1,\ldots,t_n)^{\top}$ and all the corresponding covariates in the
random-effects as $\bs{X}_n = (\bs{x}_1,\ldots,\bs{x}%
_n)^{\top}$, where $\bs{x}_i\in \mathcal{X}\subset \mathbb{R}^{p_x}$
are independently drawn from a distribution $\mathcal{U}(\bs{x})$.
Let $p(\bs{y}_n|\tau_0,\bs{X}_n)$ be the density function of
$\bs{y}_n$ given $\tau_0$ and $\bs{X}_n$, where $%
\tau_{0}(\cdot)$ is the true underling function and hence the true mean of $%
y_i$ is given by $\mu(t_i)+\tau_0(\bs{x}_i)$. Let
\begin{equation*}
p_{\bs{\theta}}(\bs{y}_n|\bs{X}_n)=\int_{\mathcal{F}%
} p(\bs{y}_n|\tau,\bs{X}_n)~\tup{d}p_{\bs{\theta}%
}(\tau)
\end{equation*}
be the density of $\bs{y}_n$ given $\bs{X}_n$ based on the
HPFR model, where $p_{\bs{\theta}}(\tau)$ is a measure of the
stochastic process $\tau(\cdot)$ on space $\mathcal{F}=\{\tau(\cdot):~%
\mathcal{X}\rightarrow \mathbb{R}\}$, and $\bs{\theta}$ contains all
the parameters in the covariance function of $\tau(\cdot)$. Denote $\tup{%
D}[p_1, p_2]=\int(\log p_1 - \log p_2)~\tup{d}p_1$ as the
Kullback-Leibler divergence between $p_1$ and $p_2$, we have the following
result.

\begin{theorem}
\label{consistency} Under appropriate conditions:
\begin{description}
  \item[(C1)] The covariance kernel function $C(\cdot,\cdot;\bs{\theta})$ is continuous in $\bs{\theta}$ and $\widehat{\bs{\theta}}\rightarrow \bs{\theta}$ almost surely as $n \rightarrow \infty$;
  \item[(C2)] The reproducing kernel Hilbert space (RKHS) norm $\|\tau_0\|_c$ is bounded;
  \item[(C3)] The expected regret term $\tup{E}_{\bs{X}_n}(\log |\bs{I}_n+\phi^{-1}\bs{C}_n|)=o(n)$,
\end{description}
 we have
\begin{equation}
\frac{1}{n}\tup{E}_{\bs{X}_{n}}(\tup{D}[p(\bs{y}%
_{n}|\tau _{0},\bs{X}_{n}),~p_{\widehat{\bs{\theta}}}(%
\bs{y}_{n}|\bs{X}_{n})])\rightarrow 0\quad \text{as}\quad
n\rightarrow \infty ,
\end{equation}
where $p_{\widehat{\bs{\theta}}}(\bs{y}_{n}|\bs{X}%
_{n})$ is the estimated density of $\bs{y}_{n}$ under the HPFR
model, $\tup{E}_{\bs{X}_{n}}$ denotes the expectation under the
distribution of $\bs{X}_{n}$.
\end{theorem}

Proof is in supplementary materials. Note here that $p(\bs{y}_{n}|\tau _{0},\bs{X}_{n})$ is the
density of the true model and $p_{\widehat{\bs{\theta}}} (\bs{y}_{n}|\bs{X}_{n})$ is that of the working model, and it
achieves information consistency, i.e., the density of working model converges to that of true model. The estimation of the mean structure $\mu_m(t)$ is based on all independent subjects and has been proved to be consistent in many functional models.\citep{Ramsay05,Yao05,Li10}
Under the working model \eqref{HPFR} of the observations, the consistency of the estimators $\widehat{\bs{\Theta}}$ is easily hold since it is essentially an MLE of a longitudinal model. The expected regret term $\tup{E}_{\bs{X}_n}(\log |\bs{I}_n+\phi^{-1}\bs{C}_n|)$ is of order $o(n)$ for some widely used covariance kernels.\citep{Seeger08}

\section{Simulation studies}

We now investigate the performance of the HPFR model in terms of accuracy
and robustness for both estimation and prediction. We will compare models
with the following three types of process: N (GP), T with $\nu =4$ and SL
with $\nu =1.3$.

The mean curve of the true model is $\mu(t)=0.8\sin ((0.5t)^3)$ with $t$
equally distributed in $[-4, 4]$. The random terms $\widetilde{ \tau}_m(t)$%
's are generated under a SMGP by setting: $x_m(t)=2.5t$ with $%
\bs{\theta}^{\top}=(v_0,w)=(0.04, 1)$ for the nonlinear
random-effects, $w_m(t)=0.5t$ with $\phi_b= 0.01$ for the linear
random-effects, and $\phi_{\varepsilon}=0.01$ for the random error.

To compare the performance of different models, we generate data of twenty
independent subjects under one of the following five schemes: (I): GPFR;
(II) TPFR with $\nu =4$; (III) GPFR with curve disturbance in the 5-th
subject (increase the amplitude of $\mu _{5}(t)$ from 0.8 to 4); (IV) GPFR
with outliers in the 10-th subject (jump the region $\{y_{10}(t)|t\in
\lbrack -1,1]\}$ upward 2 units); (V) Combine III and IV. We then use three
models (N, T, SL) to fit the data. We estimate the unknown parameters and
then calculate the marginal mean curve and random terms for each subject.
The functional coefficients involved in the fixed mean term are approximated
by cubic B-splines with 18 knots equally spaced in $[-4,4]$ which is
suitable for our simulations. In practice, the number of basis functions can
be chosen by BIC or other methods.

\subsection{Estimation for population-average inferences}

We first study the performances of HPFR models for inferences about the
marginal mean $\tup{E}[y(t)]=\mu (t)$. Table~\ref{simu-tab-1} reports
the RMSE between $\mu (t)$ and $\widehat{\mu }(t)$ for $n_{m}=31$ and $61$
based on fifty replications. Under `T' and `SL' we use the $t$-process and
slash process with fixed values of $\nu $, while under `T1' and `SL1' we
estimate it as well as other parameters. The results under Scheme I show
little difference among three models when the data are generated from the
GPFR model. However, if the data are generated from the TPFR model (Scheme
II), the GPFR model gives a poor fitting. This indicates that HPFR models
provide robust fittings against a misspecification of distribution.

We added two types of outliers respectively in Schemes III and IV and both
of them in Scheme V. The results obtained from the HPFR models with
heavy-tailed process are much better than the GPFR model, indicating that
the HPFR models are robust against outliers. The performances of two HPFR
models when $\nu $ is estimated (T1 and SL1) are very close to the models
when $\nu $ fixed (T and SL). Moreover, the choice between T and SL is not
crucial, since their results are quite similar.

\setlength{\tabcolsep}{0.3em}
\begin{table}[h]
\footnotesize\sf\centering
\begin{threeparttable}[b]
\caption{RMSE between true mean curve and its estimation.\label{simu-tab-1}}
\begin{tabular}{cccccccccccc}
\toprule
\multirow{2}{*}{Scheme} & \multicolumn{5}{c}{$n_m=31$} &  &
\multicolumn{5}{c}{$n_m=61$} \\ \cline{2-6}\cline{8-12}
& N & T & T1 & SL & SL1 &  & N & T & T1 & SL & SL1 \\
\midrule
I & 0.055 & 0.057 & 0.055 & 0.056 & 0.055 &  & 0.053 & 0.055 & 0.054 & 0.054
& 0.053 \\
II & 0.076 & 0.056 & 0.056 & 0.056 & 0.057 &  & 0.073 & 0.055 & 0.055 & 0.056
& 0.056 \\
III & 0.111 & 0.058 & 0.058 & 0.057 & 0.056 &  & 0.105 & 0.056 & 0.056 &
0.056 & 0.055 \\
IV & 0.076 & 0.059 & 0.059 & 0.059 & 0.059 &  & 0.075 & 0.056 & 0.056 & 0.056
& 0.056 \\
V & 0.117 & 0.057 & 0.057 & 0.056 & 0.055 &  & 0.116 & 0.055 & 0.055 & 0.055
& 0.055 \\
\bottomrule
\end{tabular}
\begin{tablenotes}
\item T and SL: the $t$-process and slash process with fixed degrees;
\item T1 and SL1: the $t$-process and slash process with estimated degrees.
\end{tablenotes}
\end{threeparttable}
\end{table}

Figure~4 shows the estimation of mean curve together with its
$95\%$ confidence interval under Scheme V with $n_{m}=61$ when we use the
GPFR and TPFR model from one simulated data set. It shows clearly that TPFR
achieves a smaller bias and narrow but precise confidence intervals for the
marginal means. Asymptotic normality of $\widehat{\mu }(t)$ is well
established, so that these confidence intervals will be asymptotically
correct.

\begin{figure}[h!]
	\centering
	\fbox{%
		\includegraphics[width=9cm,height=4cm]{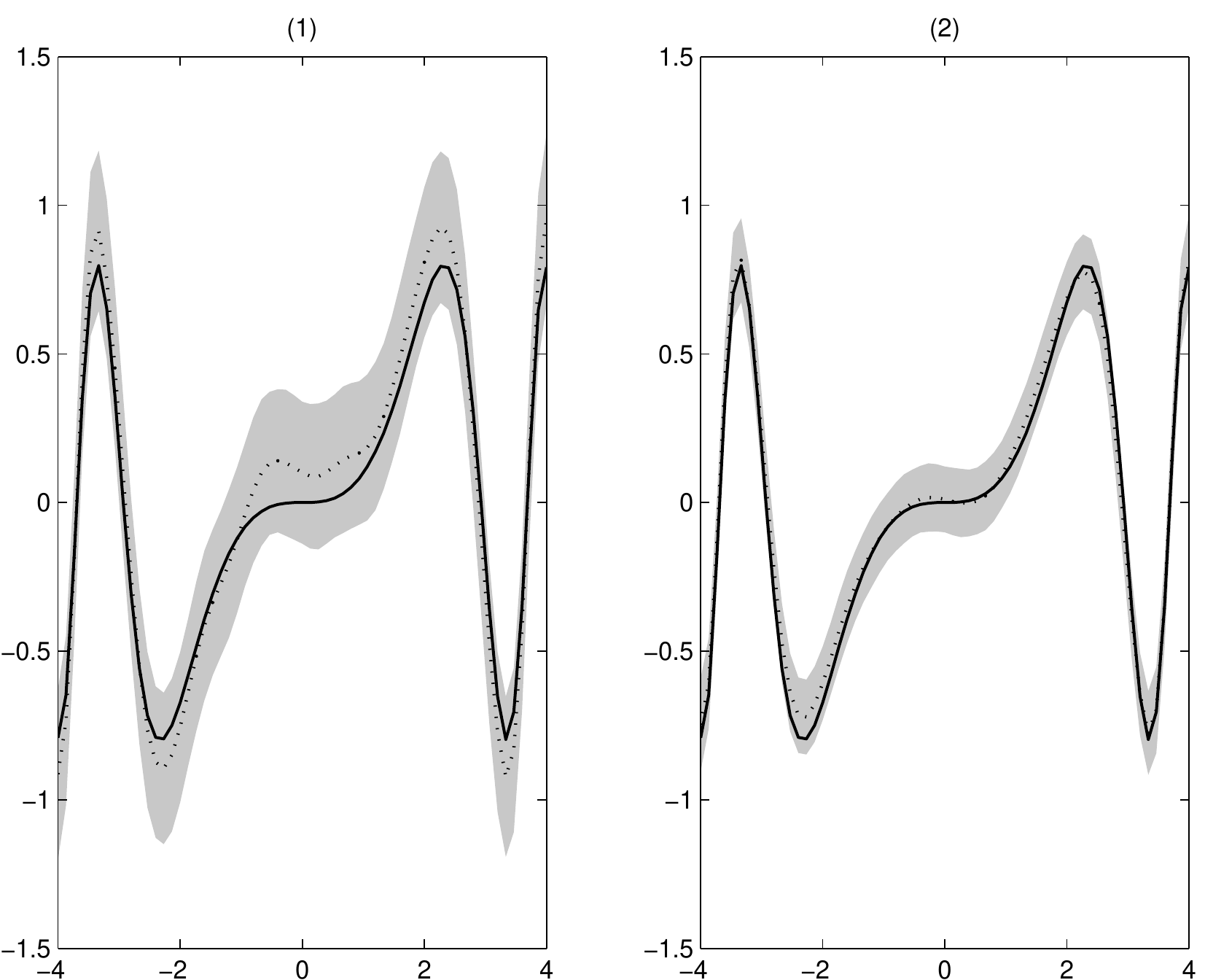}
	}
	\caption{{\protect\scriptsize Estimation of mean curve under Scheme V by
			using: (1) GPFR; (2) TPFR. Solid line: the true mean curve; dotted line: the
			estimation of the mean curve; shaded area: 95\% confidence interval.}}
	\label{simu-fig-1}
\end{figure}

\subsection{Prediction for subject-specific inferences}

We now study inferences about conditional mean, which involves prediction of
random-effects $\tau _{m}(t)$ under a conditional model \eqref{model}.\citep{Lee04}
Table~\ref{simu-tab-2} lists the RMSE between true
random-effects $\tau _{m}(t)$ and their predictions $\widehat{\tau }_{m}(t)$
for all subjects under Schemes I and II. Under Schemes III-V, we calculate
the RMSE for the subjects except the one with outliers. The RMSEs become
smaller for all cases as $n_{m}$ increases. This is because the information
about the random-effects is mainly provided by each individual subject and
the accuracy is therefore dependent mainly on the sample size of each
individual subject. The two HPFR models outperform the GPFR model when the
data come from TPFR or GPFR with outliers, which is consistent with the
previous findings.

\setlength{\tabcolsep}{0.8em}
\begin{table}[h]
\footnotesize\sf\centering
\caption{RMSE between true random terms and their predictions.\label{simu-tab-2}}
\begin{tabular}{cccccccc}
\toprule
\multirow{2}{*}{Scheme} & \multicolumn{3}{c}{$n_m=31$} &  &
\multicolumn{3}{c}{$n_m=61$} \\ \cline{2-4}\cline{6-8}
& N & T & SL &  & N & T & SL \\
\midrule
I & 0.084 & 0.084 & 0.084 &  & 0.073 & 0.073 & 0.073 \\
II & 0.120 & 0.113 & 0.113 &  & 0.105 & 0.092 & 0.092 \\
III & 0.128 & 0.084 & 0.084 &  & 0.116 & 0.073 & 0.072 \\
IV & 0.097 & 0.087 & 0.086 &  & 0.086 & 0.073 & 0.073 \\
V & 0.140 & 0.087 & 0.087 &  & 0.126 & 0.075 & 0.075 \\
\bottomrule
\end{tabular}
\end{table}

It is important to make interval statements about prediction of individual
subject. PIs based on normal assumption with the moments \eqref{conditional expect} and
\eqref{conditional
variance} is called `PL0' (plug-in method with a normal approximation), and
those based on quantiles from the predictive distribution $p_{\widehat{%
\bs{\Theta}}}(\tau _{m}(t)|\bs{\mathcal{D}})$ is called
`PL1' (plug-in method with the predictive distribution). In this paper, we
propose the bootstrap (BTS) PIs in Section 4.3.

We study the PIs for $\tau _{m}(t)$. Table~\ref{simu-tab-3} shows the
coverage probabilities (CPs) of pointwise PIs for the random terms, and
Table~\ref{simu-tab-4} shows their average lengths. For BTS, in each
replication, we generate 1000 bootstrap samples from the predictive
distribution. The value of $J$ in \eqref{predictive
distribution} is set as 50, which is big enough to provide a reasonably
accurate value to approximate the integration.

\setlength{\tabcolsep}{0.4em}
\begin{table}[h]
\footnotesize\sf\centering
\begin{threeparttable}[b]
\caption{CPs (\%) of the PIs for random terms.\label{simu-tab-3}}
\begin{tabular}{cccccccccccccc}
\toprule
\multirow{2}{*}{$n_m$} & \multirow{2}{*}{NCL (\%)} & \multirow{2}{*}{Scheme}
& \multicolumn{3}{c}{PL0} &  & \multicolumn{3}{c}{PL1} &  &
\multicolumn{3}{c}{BTS} \\ \cline{4-6}\cline{8-10}\cline{12-14}
&  &  & N & T & SL &  & N & T & SL &  & N & T & SL \\
\midrule
\multirow{9}{*}{31} & \multirow{3}{*}{80} & I & 70.5 & 71.2 & 71.1 &  & 70.3
& 70.6 & 70.5 &  & 79.4 & 80.0 & 81.5 \\
&  & II & 73.9 & 70.0 & 69.9 &  & 73.8 & 69.4 & 69.4 &  & 82.2 & 78.5 & 78.1
\\
&  & V & 56.0 & 68.4 & 68.5 &  & 55.9 & 67.8 & 68.0 &  & 83.6 & 79.3 & 80.5
\\ \cline{2-14}
& \multirow{3}{*}{90} & I & 82.1 & 82.4 & 82.5 &  & 81.9 & 82.2 & 82.2 &  &
89.4 & 89.8 & 90.9 \\
&  & II & 83.9 & 81.5 & 81.5 &  & 83.8 & 81.2 & 81.3 &  & 90.6 & 88.9 & 88.8
\\
&  & V & 69.3 & 80.0 & 80.0 &  & 69.1 & 79.8 & 79.9 &  & 93.6 & 89.6 & 90.4
\\ \cline{2-14}
& \multirow{3}{*}{95} & I & 88.9 & 88.9 & 88.9 &  & 88.8 & 89.0 & 89.0 &  &
94.5 & 94.8 & 95.7 \\
&  & II & 89.9 & 88.0 & 88.3 &  & 89.8 & 88.2 & 88.4 &  & 94.5 & 94.2 & 94.1
\\
&  & V & 78.8 & 87.3 & 87.4 &  & 78.6 & 87.4 & 87.5 &  & 97.5 & 94.7 & 95.2
\\ \hline
\multirow{9}{*}{61} & \multirow{3}{*}{80} & I & 64.2 & 64.5 & 64.5 &  & 64.2
& 64.3 & 64.2 &  & 78.7 & 78.7 & 81.8 \\
&  & II & 66.5 & 65.4 & 65.8 &  & 66.4 & 65.1 & 65.5 &  & 80.4 & 79.1 & 78.1
\\
&  & V & 45.4 & 62.5 & 62.7 &  & 45.3 & 62.2 & 62.4 &  & 83.7 & 80.3 & 82.4
\\ \cline{2-14}
& \multirow{3}{*}{90} & I & 76.4 & 76.5 & 76.6 &  & 76.3 & 76.4 & 76.4 &  &
89.0 & 88.8 & 91.4 \\
&  & II & 77.8 & 77.3 & 77.5 &  & 77.8 & 77.1 & 77.4 &  & 89.5 & 89.3 & 88.5
\\
&  & V & 57.4 & 74.4 & 74.5 &  & 57.4 & 74.2 & 74.3 &  & 93.4 & 90.2 & 91.5
\\ \cline{2-14}
& \multirow{3}{*}{95} & I & 84.1 & 84.2 & 84.2 &  & 83.9 & 84.2 & 84.2 &  &
94.1 & 94.2 & 95.9 \\
&  & II & 84.9 & 84.9 & 85.1 &  & 84.8 & 84.9 & 85.1 &  & 94.0 & 94.5 & 94.0
\\
&  & V & 67.0 & 82.2 & 82.3 &  & 67.0 & 82.1 & 82.4 &  & 97.4 & 95.0 & 95.8
\\
\bottomrule
\end{tabular}
\begin{tablenotes}
\item CP: coverage probabilities; PI: prediction interval; NCL: nominal confidence level;
\item PL0: plug-in method with a normal approximation;
\item PL1: (plug-in method with the predictive distribution;
\item BTS: bootstrap prediction.
\end{tablenotes}
\end{threeparttable}
\end{table}

\setlength{\tabcolsep}{0.4em}
\begin{table}[h]
\scriptsize\sf\centering
\caption{Lengths of the PIs for random terms.\label{simu-tab-4}}
\begin{tabular}{cccccccccccccc}
\toprule
\multirow{2}{*}{$n_m$} & \multirow{2}{*}{NCL (\%)} & \multirow{2}{*}{Scheme}
& \multicolumn{3}{c}{PL0} &  & \multicolumn{3}{c}{PL1} &  &
\multicolumn{3}{c}{BTS} \\ \cline{4-6}\cline{8-10}\cline{12-14}
&  &  & N & T & SL &  & N & T & SL &  & N & T & SL \\
\midrule
\multirow{9}{*}{31} & \multirow{3}{*}{80} & I & 0.176 & 0.179 & 0.178 &  &
0.176 & 0.177 & 0.176 &  & 0.214 & 0.218 & 0.226 \\
&  & II & 0.246 & 0.225 & 0.225 &  & 0.246 & 0.223 & 0.223 &  & 0.301 & 0.263
& 0.261 \\
&  & V & 0.234 & 0.177 & 0.176 &  & 0.233 & 0.175 & 0.174 &  & 0.398 & 0.223
& 0.227 \\ \cline{2-14}
& \multirow{3}{*}{90} & I & 0.225 & 0.230 & 0.228 &  & 0.225 & 0.229 & 0.228
&  & 0.275 & 0.281 & 0.291 \\
&  & II & 0.316 & 0.289 & 0.289 &  & 0.316 & 0.288 & 0.288 &  & 0.386 & 0.339
& 0.336 \\
&  & V & 0.300 & 0.228 & 0.226 &  & 0.300 & 0.227 & 0.225 &  & 0.510 & 0.287
& 0.292 \\ \cline{2-14}
& \multirow{3}{*}{95} & I & 0.269 & 0.274 & 0.272 &  & 0.268 & 0.275 & 0.273
&  & 0.328 & 0.336 & 0.347 \\
&  & II & 0.376 & 0.344 & 0.344 &  & 0.376 & 0.346 & 0.346 &  & 0.460 & 0.405
& 0.402 \\
&  & V & 0.357 & 0.271 & 0.269 &  & 0.357 & 0.273 & 0.270 &  & 0.605 & 0.343
& 0.349 \\ \hline
\multirow{9}{*}{61} & \multirow{3}{*}{80} & I & 0.134 & 0.136 & 0.135 &  &
0.134 & 0.135 & 0.135 &  & 0.182 & 0.184 & 0.197 \\
&  & II & 0.187 & 0.170 & 0.170 &  & 0.187 & 0.169 & 0.169 &  & 0.256 & 0.216
& 0.212 \\
&  & V & 0.169 & 0.134 & 0.134 &  & 0.169 & 0.133 & 0.133 &  & 0.360 & 0.196
& 0.206 \\ \cline{2-14}
& \multirow{3}{*}{90} & I & 0.172 & 0.174 & 0.174 &  & 0.172 & 0.174 & 0.174
&  & 0.234 & 0.236 & 0.253 \\
&  & II & 0.240 & 0.218 & 0.218 &  & 0.240 & 0.218 & 0.217 &  & 0.328 & 0.278
& 0.272 \\
&  & V & 0.217 & 0.172 & 0.172 &  & 0.217 & 0.172 & 0.171 &  & 0.462 & 0.252
& 0.265 \\ \cline{2-14}
& \multirow{3}{*}{95} & I & 0.205 & 0.208 & 0.207 &  & 0.205 & 0.208 & 0.208
&  & 0.278 & 0.281 & 0.301 \\
&  & II & 0.286 & 0.260 & 0.259 &  & 0.286 & 0.260 & 0.260 &  & 0.390 & 0.332
& 0.324 \\
&  & V & 0.258 & 0.205 & 0.204 &  & 0.258 & 0.206 & 0.205 &  & 0.549 & 0.300
& 0.315 \\
\bottomrule
\end{tabular}
\end{table}

Note that results between PL0 and PL1 are quite similar because a normal
approximation may work well in these cases. However, the CPs for both of
them are smaller than the nominal confidence levels (NCLs) since they do not
take account of the uncertainty in estimating the parameters. We see that
the BTS overcomes this drawback and maintains NCLs. When the data are
generated from a GPFR model under Scheme I, all three models provide good
results and the difference is ignorable. Under Scheme II the data are
generated from TPFR model. The advantage of HPFR models is reflected in the
precise CPs for both models but have a tighter PI than GPFR (see lengths of
PIs of BTS methods from GPFR and HPFR models in Table~\ref{simu-tab-4}). The
data under Scheme II and V are respectively a distribution misspecification
and the presence of outliers. We can see the reduction of interval length
using HPFR is much bigger under the presence of outliers than a distribution
misspecification. The PI becomes wider and the CP is higher than the nominal
levels under the GPFR model, while the two HPFR models maintain the
robustness advantage reflected in the narrow PIs with precise CPs. This is
further illustrated in Figure~5 which exhibits the results
for all subjects under Scheme V when $n_{m}=61$ from one simulated data set. It
vividly shows that: the BTS-based PIs (dark grey) are slightly wider than
the PL-based PIs (light grey), and the TPFR-based PIs (even rows) are narrow
but more accurate to cover the true random terms (solid lines) compared with
GPFR-based PIs (odd rows).

\begin{figure}[h!]
	\centering
	\fbox{%
		\includegraphics[width=11cm,height=10cm]{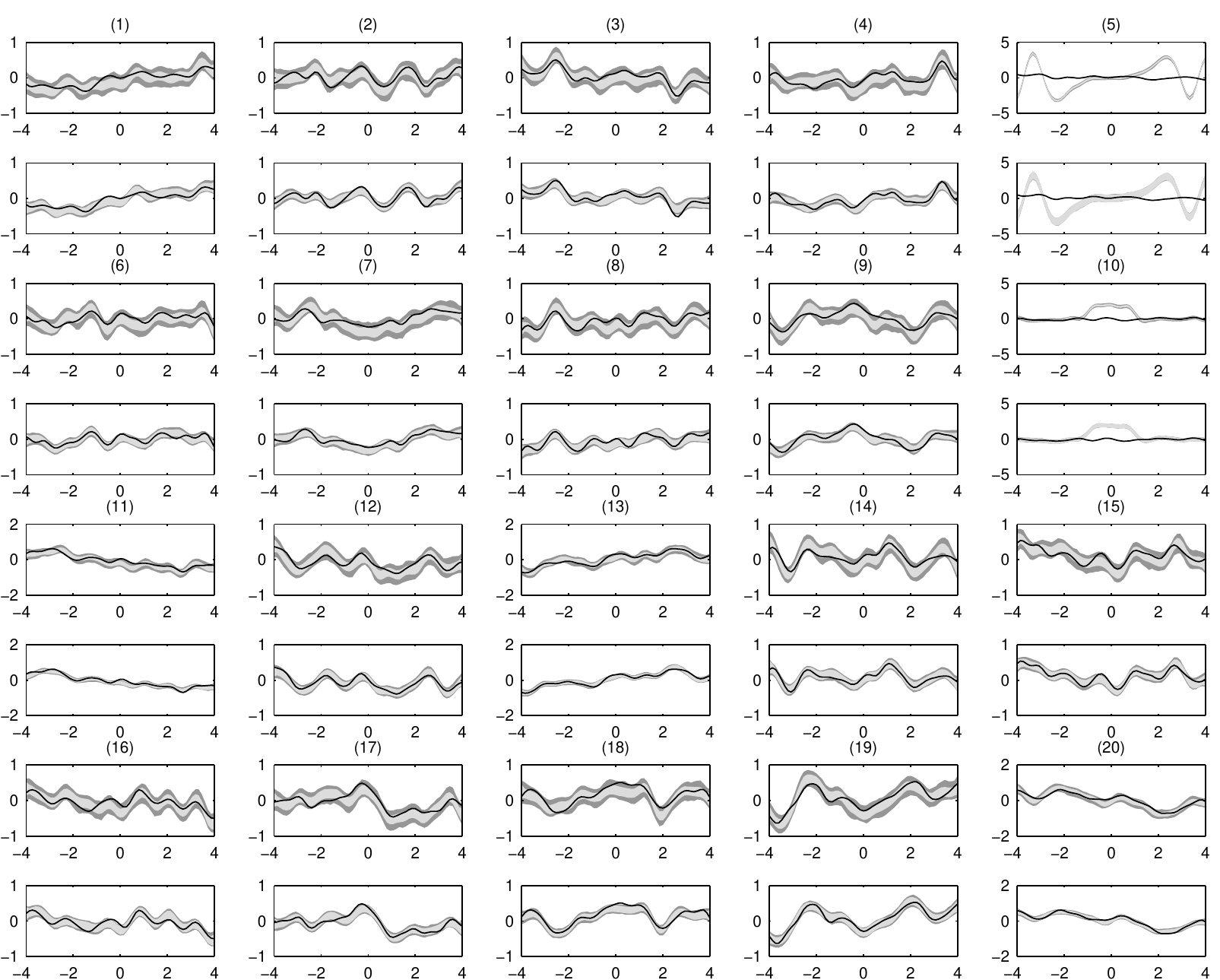}
	}
	\caption{{\protect\scriptsize Predictions of all twenty random terms under
			Scheme V. Odd rows: GPFR-based prediction; Even rows: TPFR-based prediction.
			Solid line: true underlying random term; Dark grey: $95\%$ BTS-based PI;
			Light grey: $95\%$ PL1-based PI.}}
	\label{simu-fig-2}
\end{figure}

\subsection{Prediction for a new subject}

We have studied the prediction of subject-specific curves for individuals
who are observed. Now, we assess the performance of prediction of $y_{M+1}(t)
$ for a new subject. We generate data of twenty one ($M+1$) subjects with $%
n_{m}=61$ and take the responses from the last (the 21-th) subject as our
test target. Here we consider Schemes I, II and V as described before along
with a new one: Scheme VI. Under the new scheme the data are generated from
GPFR, the same as Scheme I, but with larger random errors in the test
subject (increase $\phi_{\varepsilon }$ from 0.01 to 0.05). We use half (30
numbers) of the data from the test subject together with the data from other
twenty subjects as observed data, and try to predict the other half (31
numbers) responses (test data) of the test subject. We consider two types of
test data: one is randomly chosen, and the other is chosen from the second
half, i.e., all the data in $t_{i}\in [0, 4]$. They represent
respectively interpolation and extrapolation. The later is more challenging
than the former. Tables~\ref{simu-tab-5}-\ref{simu-tab-7} give in turn the
RMSEs, CPs and lengths of PIs of the BTS-based predictions calculated from
fifty replications.

The values of RMSE are reported in Table~\ref{simu-tab-5}. This can be used
to measure the performance of prediction. For interpolation, all three
models give very similar results under Schemes I, II and VI, meaning all of
them perform pretty well and are robust when the distribution is
misspecified. However, the two HPFR models perform more robustly than the
GPFR model in the presence of outliers under Scheme V. The results for
extrapolation almost tell the same story. Overall the errors for
extrapolation are larger than the errors for interpolation.

We report the values of CPs and the length of PIs in Tables~\ref{simu-tab-6}
and \ref{simu-tab-7}. We have some interesting findings here. First of all,
all three models give similar results under Scheme I and the values of CP
are all close to the nominal level. This shows good robustness for the two
HPFR models since the distribution is misspecified for those two models
under Scheme I. The CP values of the two HPFR models are still close to the
NCLs under Schemes II and V. But the performance of GPFR is very unreliable.
For example, the CPs are $88.3\%$ for interpolation and $98.5\%$ for
extrapolation when NCL is $80\%$ under Scheme V. For Scheme VI, the PIs
under GPFR are quite narrow leading to smaller CPs. For example, the CPs
under GPFR are $69.3\%$ for interpolation and $84.9\%$ for extrapolation
when NCL is $95\%$. The GPFR model is very sensitive to the possible
fluctuation in a new subject. In contrast, the two HPFR models give better
results although those two models also suffer from both distribution
specification and high fluctuation in the test subject.

\setlength{\tabcolsep}{0.8em}
\begin{table}[h]
\footnotesize\sf\centering
\begin{threeparttable}[b]
\caption{RMSE between true responses and their predictions of a new
subject.\label{simu-tab-5}}
\begin{tabular}{cccccccc}
\toprule
\multirow{2}{*}{Scheme} & \multicolumn{3}{c}{Interpolation} &  &
\multicolumn{3}{c}{Extrapolation} \\ \cline{2-4}\cline{6-8}
& N & T & SL &  & N & T & SL \\
\midrule
I & 0.138 & 0.137 & 0.138 &  & 0.242 & 0.243 & 0.243 \\
II & 0.198 & 0.197 & 0.197 &  & 0.311 & 0.305 & 0.306 \\
V & 0.146 & 0.137 & 0.137 &  & 0.277 & 0.249 & 0.249 \\
VI & 0.277 & 0.277 & 0.277 &  & 0.315 & 0.315 & 0.316 \\
\bottomrule
\end{tabular}
\begin{tablenotes}
\item Interpolation: the test data is randomly chosen from the new subject;
\item Extrapolation: the test data is chosen from the second half of the new subject.
\end{tablenotes}
\end{threeparttable}
\end{table}

\setlength{\tabcolsep}{0.8em}
\begin{table}[h]
\footnotesize\sf\centering
\caption{CPs (\%) of the PIs for responses of a new subject.\label{simu-tab-6}}
\begin{tabular}{ccccccccc}
\toprule
\multirow{2}{*}{NCL (\%)} & \multirow{2}{*}{Scheme} & \multicolumn{3}{c}{
Interpolation} &  & \multicolumn{3}{c}{Extrapolation} \\
\cline{3-5}\cline{7-9}
&  & N & T & SL &  & N & T & SL \\
\midrule
\multirow{4}{*}{80} & I & 81.1 & 80.6 & 82.1 &  & 76.4 & 77.0 & 78.4 \\
& II & 85.9 & 79.6 & 79.9 &  & 82.5 & 80.6 & 79.8 \\
& V & \textbf{88.3} & 79.5 & 80.7 &  & \textbf{98.5} & 79.7 & 80.4 \\
& VI & 49.9 & 73.9 & 74.2 &  & 65.6 & 90.5 & 90.4 \\ \hline
\multirow{4}{*}{90} & I & 90.0 & 90.3 & 90.6 &  & 87.5 & 88.1 & 88.5 \\
& II & 92.3 & 90.8 & 90.7 &  & 91.0 & 91.2 & 90.5 \\
& V & 96.1 & 90.2 & 90.8 &  & 99.7 & 90.4 & 91.0 \\
& VI & 60.7 & 84.8 & 85.4 &  & 76.8 & 96.5 & 96.6 \\ \hline
\multirow{4}{*}{95} & I & 94.7 & 94.5 & 95.5 &  & 92.8 & 94.0 & 94.1 \\
& II & 94.9 & 94.9 & 95.0 &  & 94.8 & 95.7 & 95.9 \\
& V & 98.6 & 94.6 & 95.5 &  & 99.9 & 94.8 & 95.7 \\
& VI & \textbf{69.3} & 91.2 & 91.0 &  & \textbf{84.9} & 98.5 & 98.5 \\
\bottomrule
\end{tabular}
\end{table}

\setlength{\tabcolsep}{0.8em}
\begin{table}[h]
\footnotesize\sf\centering
\caption{Lengths the PIs for responses of a new subject.\label{simu-tab-7}}
\begin{tabular}{ccccccccc}
\toprule
\multirow{2}{*}{NCL (\%)} & \multirow{2}{*}{Scheme} & \multicolumn{3}{c}{
Interpolation} &  & \multicolumn{3}{c}{Extrapolation} \\
\cline{3-5}\cline{7-9}
&  & N & T & SL &  & N & T & SL \\
\midrule
\multirow{4}{*}{80} & I & 0.353 & 0.356 & 0.365 &  & 0.584 & 0.598 & 0.606
\\
& II & 0.494 & 0.429 & 0.428 &  & 0.787 & 0.735 & 0.729 \\
& V & 0.487 & 0.353 & 0.359 &  & 1.390 & 0.642 & 0.650 \\
& VI & 0.365 & 0.623 & 0.624 &  & 0.589 & 1.100 & 1.108 \\ \hline
\multirow{4}{*}{90} & I & 0.453 & 0.461 & 0.471 &  & 0.750 & 0.774 & 0.784
\\
& II & 0.634 & 0.555 & 0.553 &  & 1.010 & 0.950 & 0.943 \\
& V & 0.625 & 0.456 & 0.463 &  & 1.787 & 0.831 & 0.840 \\
& VI & 0.468 & 0.806 & 0.809 &  & 0.755 & 1.423 & 1.435 \\ \hline
\multirow{4}{*}{95} & I & 0.541 & 0.554 & 0.565 &  & 0.894 & 0.929 & 0.941
\\
& II & 0.756 & 0.667 & 0.665 &  & 1.203 & 1.142 & 1.133 \\
& V & 0.746 & 0.548 & 0.555 &  & 2.131 & 0.998 & 1.007 \\
& VI & 0.558 & 0.969 & 0.973 &  & 0.900 & 1.710 & 1.726 \\
\bottomrule
\end{tabular}
\end{table}

The predictions with
a $95\%$ NCL from one simulated data set are plotted in Figure~6
under Scheme V (upper panel) and VI (lower panel) when $n_{m}=61
$. For Scheme V, the PIs under the TPFR model (light grey) are narrow than
those under the GPFR model (dark grey), which is more significant for
extrapolation (upper-right). Bear in mind that the cover probability for the
former is more close to the NCL $95\%$ than the latter. For Scheme VI, the
PIs under the GPFR model are too narrow to cover a large proportion of
observations, leading to a very small cover probability.

\begin{figure}[h!]
	\centering
	\fbox{%
		\includegraphics[width=9cm,height=7cm]{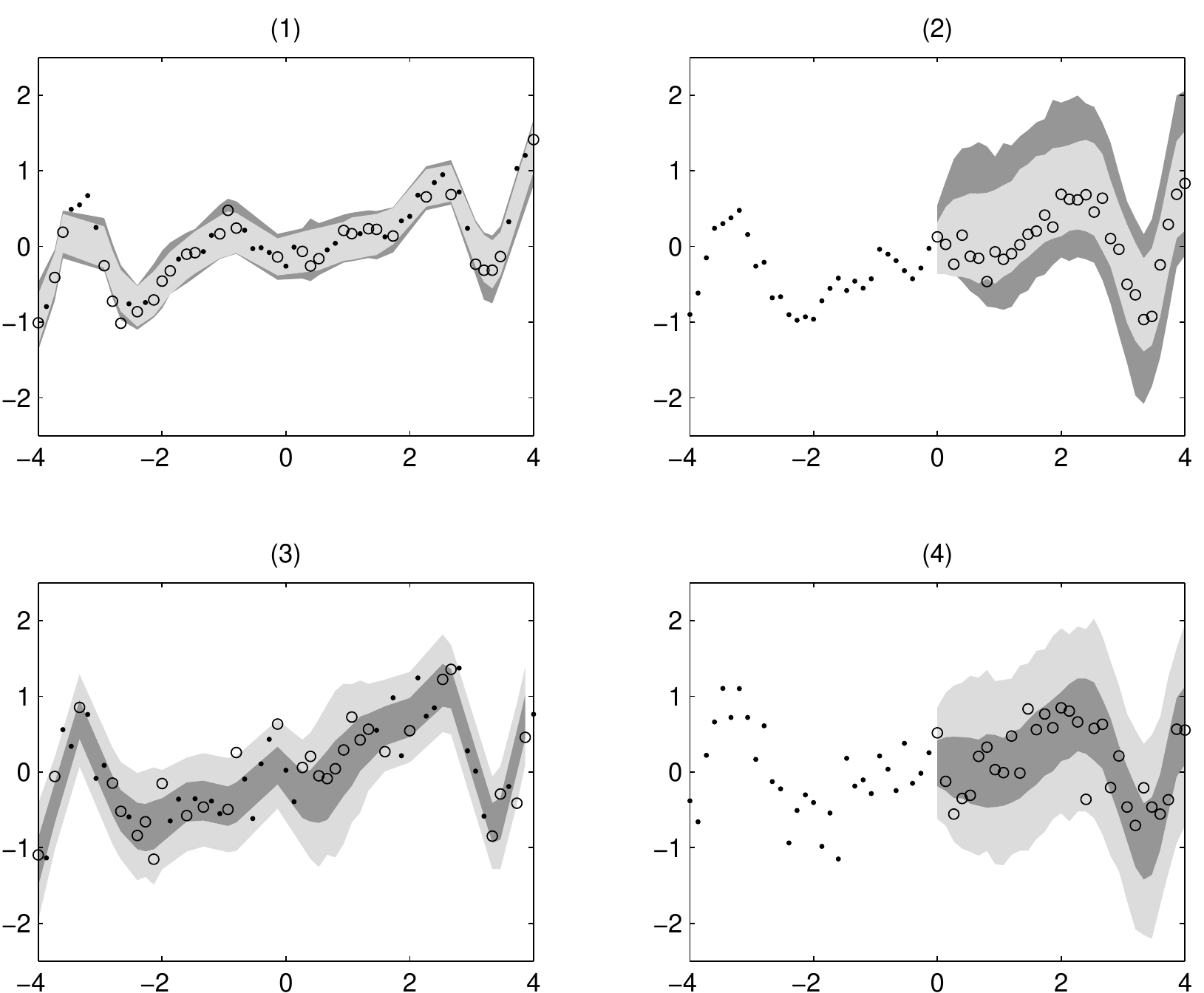} }
	\caption{{\protect\scriptsize Bootstrap prediction for a new subject.
			Upper-left: Interpolation under Scheme V; Upper-right: Extrapolation under
			Scheme VI; Lower-left: Interpolation under Scheme VI; Lower-right:
			Extrapolation under Scheme VI. Dark grey: $95\%$ BTS-based PI under the GPFR
			model; Light grey: $95\%$ BTS-based PI under the TPFR model. Dot: true
			observed response; Circle: true testing response.}}
	\label{simu-fig-3}
\end{figure}

Overall, we show that the HPFR models perform better than the GPFR model if
the distribution is misspecified and if there are outliers. The proposed BTS
based-PIs under HPFR models are quite effective for subject-specific
predictions.

\section{Concluding remarks and discussion}

GP models are widely used for the analysis of medical data. The GPFR model can describe subject-specific characteristics nonparametrically by using a GP via a flexible covariance structure, and can cope with multidimensional covariates in a nonparametric way. However, as we demonstrated in simulation studies, it is sensitive to distribution misspecification and outliers. To overcome this drawback we proposed the HPFR model. We have shown via a comprehensive simulation study that the proposed model has a good property of robustness, giving accurate results when the distribution is misspecified or when the data are contaminated by outliers. The PIs for the subject-specific curves have not been well developed. We propose the use of bootstrap PIs. The simulation study shows that the proposed bootstrap PIs improve existing PIs with respect to the CP and the length of the PI. Computing code of Matlab or R can be provided upon request.

The model has a large flexibility since it includes a class of process regression model for example, the one with a GP,\citep{Shi12} the T-process regression model,\citep{Wang16} and the model with slash process, mixtures of GP, exponential power process, etc.
Even though limited, according to our experience, we prefer the HP model to the GP model because the HP model improves the GP model a lot in the presence of outliers. In complex or big data, identification of all outliers would be difficult. It is reasonable to use the HP model over the GP model since it can bring robust inference and does not lose much efficiency even the data truly comes from GP.
As we demonstrated, some criteria, such as the values of BIC, predictive RMSE, standard errors of the estimators and relative change ratios by deleting potential outliers, can be used to select a specific HP model.

All computations were carried out in Matlab 2015b using a 2.4 GHz Inter i5 processor with 8.0 GB RAM. The EM algorithm is very efficient because the E-step is straightforward in our HPFR model. The computation time to achieve convergence is about 12 seconds for each replication of the simulation (Scheme V and $n_m = 61$) and 6 seconds for the example in Section 3 under the T-process model.
However, we can face with the problems from high-dimensional covariates or high frequency data. For high-dimensional covariates, penalized likelihood framework will helpful to select the crucial functional covariates.\citep{Yi11} For high frequency data, the HPFR model suffers from massive computation of inverse covariance matrices. A variety of numerical methods, such as Nystr\"{o}m method, active set and filtering approach \citep{Shi11} may be applied to solve the problem.

One crucial issue in FDA is the modelling of mean function and covariance structure.\citep{Guo02,Antoniadis07} For the mean function $\mu_m(t)$, we used nonparametric function $\bs{u}_m^{\top}\bs{\beta}(t)$ combined with a parametric (linear) function $\bs{v}_m^{\top}(t)\bs{\gamma}$. Other mean structures can also be considered. The functional coefficients $\bs{\beta}(t)$ were approximated by cubic B-splines basis functions under equally spaced knots. It is known that the number of knots tunes the bias and variance of resulting estimator. Thus, the choice of the number of knots is important for the performance of B-spline approximation. The guidance on this issue can be seen in \citet{Wu06}. In practice, the number of knots can be determined by generalized cross-validation or BIC methods. However, it is time consuming for high-dimensional and big data. Some low-rank smoothing methods, such as penalized spline smoothing \citep{Ruppert03} can be used to decrease the computational burden. For the covariance structure, we combined a parametric (linear) random-effects model $\bs{w}_m^{\top}(t)\bs{b}_m$ with a nonparametric (non-linear) random-effects $\zeta_m(\bs{x}_m(t))$ by using HPs. The parametric random-effects model $\bs{w}_m^{\top}(t)\bs{b}_m$ can provide explanatory information between the response and some covariates and characterize the heterogeneity among subjects. A diagonal matrix of $\bs{\Sigma}_b$ means that a few latent variables act independently in subjects, while a non-diagonal positive matrix of $\bs{\Sigma}_b$ implies that there are correlations among the latent variables. The nonparametric process-based random-effects $\zeta_m(\bs{x}_m(t))$ can handle flexible subject-specific deviations. The covariance kernel allows us consider multi-dimensional functional covariates to catch up the nonlinear serial correlation within subject. The kernel may represent some symmetric (mirror) effects over time if we only consider time in the process. More discussion of the covariance kernel can be seen in \citet{Rasmussen06} and \citet{Shi11}.

Model diagnosis is also an important issue in FDA. Our HPFR model side-steps the problem of heterogeneity, which means that different subjects share the same location parameters, scale parameters and also the degrees. However, this assumption is doubtful when the data are collected from different groups or sources. Recently, \citet{Fang16} proposed two tests for evaluating the variance homogeneity in mixed-effects models. Their approaches may also be adapted in the FDA models. For discussion of model-checking plots of various model-misspecification, see \citet{Lee06}. We will develop a model-based method of clustering or classification as a next work if the assumption of homogeneity does not hold.

\subsection*{Acknowledgement}

The authors thank Prof. West for providing us the renal anaemia data. The authors are grateful to the Editor, the Associate Editor and two referees, whose questions and insightful comments have greatly improved the work.

The author(s) disclosed receipt of the following financial support for the research, authorship, and/or publication of this article:
Dr Cao's work was supported by the National Science Foundation of China
(Grant No. 11301278) and the MOE (Ministry of Education in China) Project of
Humanities and Social Sciences (Grant No. 13YJC910001).

Technical details on the conditional distribution properties of SMN distributions, the information matrix and the proof of information consistency are presented in an additional document, which is distributed with the paper as Supplementary Material.


\begin{thebibliography}{99}

\bibitem[Rasmussen and Williams(2006)]{Rasmussen06}
Rasmussen~CE and Williams~CKI. \textit{Gaussian Processes for Machine Learning}.
Cambridge, MA: MIT Press, 2006.

\bibitem[Shi et al.(2007)]{Shi07}
Shi~JQ, Wang~B, Murray-Smith~R, et al. Gaussian process functional
regression modelling for batch data. \textit{Biometrics} 2007; 63(3): 714-723.

\bibitem[Shi et al.(2012)]{Shi12}
Shi~JQ, Wang~B, Will~EJ, et al. Mixed-effects GPFR models with application to
dose-response curve prediction. \textit{Stat Med} 2012; 31(26): 3165-3177.

\bibitem[Gramacy and Lian (2012)]{Gramacy12}
Gramacy~R and Lian~H. Gaussian process single-index models as emulators for
computer experiments. \textit{Technometrics} 2012; 54(1): 30-41.

\bibitem[Wang and Shi(2014)]{Wang14}
Wang~B and Shi~JQ. Generalized Gaussian process regression model for non-Gaussian
functional Data. \textit{J AM Stat Assoc} 2014; 109(507): 1123-1133.

\bibitem[Andrews and Mallows(1974)]{Andrews74}
Andrews~DF and Mallows~CL. Scale mixtures of normal distributions. \textit{J R Soc Series B Stat Methodol}
1974; 36(1): 99-102.

\bibitem[Lachos et al.(2011)]{Lachos11}
Lachos~VH, Bandyopadhyay~D and Dey~DK. Linear and nonlinear mixed-effects models for censored HIV
viral loads using normal/independent distributions. \textit{Biometrics} 2011; 67(4); 1594-1604.

\bibitem[Meza et al.(2012)]{Meza12}
Meza~C, Osorio~F and Cruz~R. Estimation in nonlinear mixed-effects models using
heavy-tailed distributions. \textit{Stat Comput} 2012; 22(1): 121-139.

\bibitem[Cao et al.(2015)]{Cao15}
Cao~CZ, Lin~JG., Shi~JQ, et al. Multivariate measurement error models for replicated data
under heavy-tailed distributions. \textit{J Chemometr} 2015; 29(8): 457-466.

\bibitem[Blas et al.(2016)]{Blas16}
Blas~B, Bolfarine~H and Lachos~VH. Heavy tailed calibration model with Berkson measurement errors for
replicated data. \textit{Chemometr Intell Lab} 2016; 156: 21-35.

\bibitem[Zhu et al.(2011)]{Zhu11}
Zhu~H, Brown~PJ and Morris~JS. Robust, adaptive functional regression in functional
mixed model framework. \textit{J AM Stat Assoc} 2011;
106(495): 1167-1179.

\bibitem[Osorio(2016)]{Osorio16}
Osorio~F. Influence diagnostics for robust P-splines using scale mixture of normal
distributions. \textit{Ann I of Stat Math} 2016; 68(3): 589-619.

\bibitem[Lee et al.(2006)]{Lee06}
Lee~Y, Nelder~JA and Pawitan~Y. \textit{Generalized linear models with random-effects, unified
analysis via H-likelihood}. London: Chapman and Hall, 2006.

\bibitem[McCulloch and Neuhaus(2011)]{McCulloch11}
McCulloch~CE and Neuhaus~JM. Prediction of random effects in linear and
generalized linear models under model misspecification. \textit{Biometrics} 2011; 67(1):
270-279.

\bibitem[Lee and Kim(2016)]{Lee16}
Lee~Y and Kim~G. H-likelihood predictive intervals for unobservables. \textit{Int
Stat Rev} 2016; Online.

\bibitem[Diggle et al.(1996)]{Diggle96}
Diggle~PJ, Liang~KY and Zeger~SL. \textit{Analysis of longitudinal data}.
New York: Oxford Univ. Press, 1996.

\bibitem[Lee and Nelder(2004)]{Lee04}
Lee~Y and Nelder~JA. Conditional and marginal models: Another view (with discussion).
\textit{Stat Sci} 2004; 19(2): 219-238.

\bibitem[Ramsay and Silverman(2005)]{Ramsay05}
Ramsay~JO and Silverman~BW. \textit{Functional data analysis}, 2nd~edn. New York: Springer, 2005.

\bibitem[Shi and Choi(2011)]{Shi11}
Shi~JQ and Choi~T. \textit{Gaussian process regression analysis for functional data}. London: Chapman
and Hall, 2011.

\bibitem[Lange and Sinsheimer(1993)]{Lange93}
Lange~K and Sinsheimer~JS. Normal/independent distributions and their
applications in robust regression. \textit{J Comput Graph Stat} 1993; 2(2): 175-198.

\bibitem[Pinheiro et al.(2001)]{Pinheiro01}
Pinheiro~JC, Liu~C and Wu~YN. Efficient algorithms for robust estimation in
linear mixed-effect models using the multivariate $t$ distribution. \textit{J Comput Graph Stat} 2001; 10(2): 249-276.

\bibitem[Savalli et al.(2006)]{Savalli06}
Savalli~C, Paula~GA and Cysneiros~FJA. Assessment of variance components in
elliptical linear mixed models. \textit{Stat Model} 2006; 6(1): 59-76.

\bibitem[Wang et al.(2016)]{Wang16}
Wang~Z, Shi~JQ and Lee~Y. Extended T-process regression models. 2016; arXiv: 1511.03402.

\bibitem[West et al.(2007)]{West07}
West~RM, Harris~K, Gilthorpe~MS, et al. Functional data analysis
applied to a randomized controlled clinical trial in hemodialysis patients
describes the variability of patient responses in the control of renal
anemia. \textit{J AM Soc Nephrol} 2007; 18(8): 2371-2376.

\bibitem[Tolman et al.(2005)]{Tolman05}
Tolman~C, Richardson~D, Bartlett~C, et al. Structured conversion from
thrice weekly to weekly erythropoietic regimens using a computerized
decision-support system: a randomized clinical study. \textit{J AM Soc Nephrol} 2005; 16(5): 1463-1470.

\bibitem[Will et al.(2007)]{Will07}
Will~EJ, Richardson~D, Tolman~C, et al. Development and exploitation of a
clinical decision support system for the management of renal anaemia. \textit{
Nephrology, Dialysis and Transplantation} 2007; 22(suppl 4): iv31-iv36.

\bibitem[Schwarz(1978)]{Schwarz78}
Schwarz~G. Estimating the dimension of a model. \textit{Ann Stat} 1978; 6(2): 461-464.

\bibitem[Fang et al.(1990)]{Fang90}
Fang~ KT, Kotz~S and Ng~KW. \textit{Symmetrical multivariate and related distributions}. London: Chapman and
Hall, 1990.

\bibitem[Lee and Nelder(1996)]{Lee96}
Lee~Y, Nelder~JA. Hierarchical Generalized Linear Models (with discussion).
\textit{J R Soc Series B Stat Methodol} 1996; 58(4): 619-678.

\bibitem[Meng and Rubin(1993)]{Meng93}
Meng~XL and Rubin~DB. Maximum likelihood estimation via the ECM algorithm: A general
framework. \textit{Biometrika} 1993; 80(2): 267-278.

\bibitem[Liu and Rubin(1994)]{Liu94}
Liu~CH and Rubin~DB. The ECME algorithm: a simple extension of EM and ECM with faster
monotone convergence. \textit{Biometrika} 1994; 81(4): 633-648.

\bibitem[Lange et al.(1989)]{Lange89}
Lange~K, Little~R and Taylor~J. Robust statistical modeling using the $t$ distribution.
\textit{J AM Stat Assoc} 1989; 84(408): 881-896.

\bibitem[Bj{\o}rnstad(1990)]{Bjornstad90}
Bj{\o}rnstad~JF. Predictive likelihood: A review. \textit{Stat Sci} 1990; 5(1): 242-265.

\bibitem[Bj{\o}rnstad(1996)]{Bjornstad96}
Bj{\o}rnstad~JF. On the generalization of the likelihood function and likelihood
principle. \textit{J AM Stat Assoc} 1996; 91(434): 791-806.

\bibitem[Seeger et al.(2008)]{Seeger08}
Seeger~MW, Kakade~SM and Foster~DP. Information consistency of nonparametric
Gaussian process methods. \textit{IEEE T Inform Theory} 2008; 54(5): 2376-2382.

\bibitem[Yao et al.(2005)]{Yao05}
Yao~F, M\"{u}ller~HG and Wang~JL. Functional data analysis for sparse longitudinal data. \textit{J AM Stat Assoc} 2005; 100(470): 577-590.

\bibitem[Li and Hsing(2010)]{Li10}
Li~Y and Hsing~T. Uniform convergence rates for nonparametric regression and principal component analysis in functional/longitudinal data. \textit{Ann Stat} 2010; 38(6): 3321-3351.

\bibitem[Yi et al.(2011)]{Yi11}
Yi~G, Shi~JQ and Choi~T. Penalized gaussian process regression and classification for high-dimensional nonlinear data. \textit{Biometrics} 2011; 67(4): 1285-1294.

\bibitem[Guo(2002)]{Guo02}
Guo~W. Functional mixed effects models. \textit{Biometrics} 2002; 58(1): 121-128.

\bibitem[Antoniadis and Sapatinas(2007)]{Antoniadis07}
Antoniadis~A and Sapatinas~T. Estimation and inference in functional mixed-effects models. \textit{Comput Stat Data Anal} 2007; 51(10): 4793-4813.

\bibitem[Wu and Zhang(2006)]{Wu06}
Wu~H and Zhang~JT. \textit{Nonparametric regression methods for longitudinal data analysis: mixed-effects modeling approaches}. New Jersey:
Wiley, 2006.

\bibitem[Ruppert et al.(2003)]{Ruppert03}
Ruppert~D, Wand~MP and Carroll~RJ. \textit{Semiparametric regression}. Cambridge: Cambridge University Press, 2003.

\bibitem[Fang et al.(2016)]{Fang16}
Fang~X, Li~J, et al. Detecting the violation of variance homogeneity in mixed models. \textit{Stat Methods Med Res}. 2016; 25(6): 2506-2520.
\end{thebibliography}
\end{document}